\documentclass[pra,aps,onecolumn,twoside,superscriptaddress]{revtex4}

\usepackage{amsmath,amsfonts,amssymb,caption,color,epsfig,graphics,graphicx,hyperref,latexsym,mathrsfs,revsymb,theorem,url,verbatim,enumerate,epstopdf,tikz,float,multirow,booktabs,appendix}
\hypersetup{colorlinks,linkcolor={blue},citecolor={red},urlcolor={blue}}
\usetikzlibrary{arrows, decorations.markings}

\tikzstyle{vecArrow} = [thick, decoration={markings,mark=at position
	1 with {\arrow[semithick]{open triangle 60}}},
double distance=1.4pt, shorten >= 5.5pt,
preaction = {decorate},
postaction = {draw,line width=1.4pt, white,shorten >= 4.5pt}]
\tikzstyle{innerWhite} = [semithick, white,line width=1.4pt, shorten >= 4.5pt]

\newtheorem{definition}{Definition}
\newtheorem{proposition}[definition]{Proposition}
\newtheorem{lemma}[definition]{Lemma}

\newtheorem{theorem}[definition]{Theorem}
\newtheorem{corollary}[definition]{Corollary}
\newtheorem{conjecture}[definition]{Conjecture}

\newtheorem{remark}[definition]{Remark}
\newtheorem{example}[definition]{Example}
\newtheorem{question}[definition]{Question}
\newtheorem{construction}{Construction}

\def\bcj{\begin{conjecture}}
\def\ecj{\end{conjecture}}
\def\bcr{\begin{corollary}}
\def\ecr{\end{corollary}}
\def\bd{\begin{definition}}
\def\ed{\end{definition}}
\def\bea{\begin{eqnarray}}
\def\eea{\end{eqnarray}}
\def\bem{\begin{enumerate}}
\def\eem{\end{enumerate}}
\def\bex{\begin{example}}
\def\eex{\end{example}}
\def\bim{\begin{itemize}}
\def\eim{\end{itemize}}
\def\bl{\begin{lemma}}
\def\el{\end{lemma}}
\def\bma{\begin{bmatrix}}
\def\ema{\end{bmatrix}}
\def\bpf{\begin{proof}}
\def\epf{\end{proof}}
\def\bpp{\begin{proposition}}
\def\epp{\end{proposition}}
\def\bqu{\begin{question}}
\def\equ{\end{question}}
\def\br{\begin{remark}}
\def\er{\end{remark}}
\def\bt{\begin{theorem}}
\def\et{\end{theorem}}

\def\squareforqed{\hbox{\rlap{$\sqcap$}$\sqcup$}}
\def\qed{\ifmmode\squareforqed\else{\unskip\nobreak\hfil
\penalty50\hskip1em\null\nobreak\hfil\squareforqed
\parfillskip=0pt\finalhyphendemerits=0\endgraf}\fi}
\def\endenv{\ifmmode\;\else{\unskip\nobreak\hfil
\penalty50\hskip1em\null\nobreak\hfil\;
\parfillskip=0pt\finalhyphendemerits=0\endgraf}\fi}
\newenvironment{proof}{\noindent \textbf{{Proof.~} }}{\qed}
\def\Dbar{\leavevmode\lower.6ex\hbox to 0pt
{\hskip-.23ex\accent"16\hss}D}
\makeatletter
\def\url@leostyle{%
  \@ifundefined{selectfont}{\def\UrlFont{\sf}}{\def\UrlFont{\small\ttfamily}}}
\makeatother
\def\bcj{\begin{conjecture}}
\def\ecj{\end{conjecture}}
\def\bcr{\begin{corollary}}
\def\ecr{\end{corollary}}
\def\bd{\begin{definition}}
\def\ed{\end{definition}}
\def\bea{\begin{eqnarray}}
\def\eea{\end{eqnarray}}
\def\bem{\begin{enumerate}}
\def\eem{\end{enumerate}}
\def\bex{\begin{example}}
\def\eex{\end{example}}
\def\bim{\begin{itemize}}
\def\eim{\end{itemize}}
\def\bl{\begin{lemma}}
\def\el{\end{lemma}}
\def\bpf{\begin{proof}}
\def\epf{\end{proof}}
\def\bpp{\begin{proposition}}
\def\epp{\end{proposition}}
\def\bqu{\begin{question}}
\def\equ{\end{question}}
\def\br{\begin{remark}}
\def\er{\end{remark}}
\def\bt{\begin{theorem}}
\def\et{\end{theorem}}

\def\btb{\begin{tabular}}
\def\etb{\end{tabular}}

\newcommand{\nc}{\newcommand}

\def\ps{\psi}

 \nc{\bbA}{\mathbb{A}} \nc{\bbB}{\mathbb{B}} \nc{\bbC}{\mathbb{C}}
 \nc{\bbD}{\mathbb{D}} \nc{\bbE}{\mathbb{E}} \nc{\bbF}{\mathbb{F}}
 \nc{\bbG}{\mathbb{G}} \nc{\bbH}{\mathbb{H}} \nc{\bbI}{\mathbb{I}}
 \nc{\bbJ}{\mathbb{J}} \nc{\bbK}{\mathbb{K}} \nc{\bbL}{\mathbb{L}}
 \nc{\bbM}{\mathbb{M}} \nc{\bbN}{\mathbb{N}} \nc{\bbO}{\mathbb{O}}
 \nc{\bbP}{\mathbb{P}} \nc{\bbQ}{\mathbb{Q}} \nc{\bbR}{\mathbb{R}}
 \nc{\bbS}{\mathbb{S}} \nc{\bbT}{\mathbb{T}} \nc{\bbU}{\mathbb{U}}
 \nc{\bbV}{\mathbb{V}} \nc{\bbW}{\mathbb{W}} \nc{\bbX}{\mathbb{X}}
 \nc{\bbZ}{\mathbb{Z}}

 \nc{\bA}{{\bf A}} \nc{\bB}{{\bf B}} \nc{\bC}{{\bf C}}
 \nc{\bD}{{\bf D}} \nc{\bE}{{\bf E}} \nc{\bF}{{\bf F}}
 \nc{\bG}{{\bf G}} \nc{\bH}{{\bf H}} \nc{\bI}{{\bf I}}
 \nc{\bJ}{{\bf J}} \nc{\bK}{{\bf K}} \nc{\bL}{{\bf L}}
 \nc{\bM}{{\bf M}} \nc{\bN}{{\bf N}} \nc{\bO}{{\bf O}}
 \nc{\bP}{{\bf P}} \nc{\bQ}{{\bf Q}} \nc{\bR}{{\bf R}}
 \nc{\bS}{{\bf S}} \nc{\bT}{{\bf T}} \nc{\bU}{{\bf U}}
 \nc{\bV}{{\bf V}} \nc{\bW}{{\bf W}} \nc{\bX}{{\bf X}}
 \nc{\bZ}{{\bf Z}} \nc{\bm}{{\bf m}} \nc{\bv}{{\bf v}}
 \nc{\ba}{{\bf a}} \nc{\be}{{\bf e}} \nc{\bu}{{\bf u}}
 \nc{\brr}{{\bf r}}

\nc{\cA}{{\cal A}} \nc{\cB}{{\cal B}} \nc{\cC}{{\cal C}}
\nc{\cD}{{\cal D}} \nc{\cE}{{\cal E}} \nc{\cF}{{\cal F}}
\nc{\cG}{{\cal G}} \nc{\cH}{{\cal H}} \nc{\cI}{{\cal I}}
\nc{\cJ}{{\cal J}} \nc{\cK}{{\cal K}} \nc{\cL}{{\cal L}}
\nc{\cM}{{\cal M}} \nc{\cN}{{\cal N}} \nc{\cO}{{\cal O}}
\nc{\cP}{{\cal P}} \nc{\cQ}{{\cal Q}} \nc{\cR}{{\cal R}}
\nc{\cS}{{\cal S}} \nc{\cT}{{\cal T}} \nc{\cU}{{\cal U}}
\nc{\cV}{{\cal V}} \nc{\cW}{{\cal W}} \nc{\cX}{{\cal X}}
\nc{\cZ}{{\cal Z}}

\nc{\hA}{{\hat{A}}} \nc{\hB}{{\hat{B}}} \nc{\hC}{{\hat{C}}}
\nc{\hD}{{\hat{D}}} \nc{\hE}{{\hat{E}}} \nc{\hF}{{\hat{F}}}
\nc{\hG}{{\hat{G}}} \nc{\hH}{{\hat{H}}} \nc{\hI}{{\hat{I}}}
\nc{\hJ}{{\hat{J}}} \nc{\hK}{{\hat{K}}} \nc{\hL}{{\hat{L}}}
\nc{\hM}{{\hat{M}}} \nc{\hN}{{\hat{N}}} \nc{\hO}{{\hat{O}}}
\nc{\hP}{{\hat{P}}} \nc{\hR}{{\hat{R}}} \nc{\hS}{{\hat{S}}}
\nc{\hT}{{\hat{T}}} \nc{\hU}{{\hat{U}}} \nc{\hV}{{\hat{V}}}
\nc{\hW}{{\hat{W}}} \nc{\hX}{{\hat{X}}} \nc{\hZ}{{\hat{Z}}}

\nc{\hn}{{\hat{n}}}

\def\ghz{\mathop{\rm GHZ}}

\def\min{\mathop{\rm min}}

\def\tr{\mathop{\rm Tr}}

\def\moa{\mathop{\rm MOA}}
\def\oa{\mathop{\rm OA}}
\def\irm{\mathop{\rm IrMOA}}
\def\iro{\mathop{\rm IrOA}}
\def\hd{\mathop{\rm HD}}
\def\md{\mathop{\rm MD}}
\def\ghm{\mathop{\rm GHM}}

\newcommand{\bra}[1]{\langle#1|}
\newcommand{\ket}[1]{|#1\rangle}

\newcommand{\ketbra}[2]{|#1\rangle\!\langle#2|}
\newcommand{\braket}[2]{\langle#1|#2\rangle}

\newcommand{\fl}[2]{\lfloor\frac{#1}{#2}\rfloor}

\def\Dbar{\leavevmode\lower.6ex\hbox to 0pt
{\hskip-.23ex\accent"16\hss}D}

\begin{document}

\title{Constructions of $k$-uniform states  from mixed orthogonal arrays}



\author{Fei Shi}
\email[]{shifei@mail.ustc.edu.cn}
\affiliation{School of Cyber Security,
	University of Science and Technology of China, Hefei, 230026, People's Republic of China.}

\author{Yi Shen}
\email[]{yishen@buaa.edu.cn}
\affiliation{School of Mathematical Sciences, Beihang University, Beijing 100191, China}
\affiliation{Department of Mathematics and Statistics, Institute for Quantum Science and Technology, University of Calgary, AB, Canada T2N 1N4}

\author{Lin Chen}
\email[]{linchen@buaa.edu.cn}
\affiliation{School of Mathematical Sciences, Beihang University, Beijing 100191, China}
\affiliation{International Research Institute for Multidisciplinary Science, Beihang University, Beijing 100191, China}

\author{Xiande Zhang}
\email[]{drzhangx@ustc.edu.cn}
\affiliation{School of Mathematical Sciences,
	University of Science and Technology of China, Hefei, 230026, People's Republic of China}

\begin{abstract}
We study $k$-uniform states in heterogeneous systems whose local dimensions are mixed. Based on the connections between  mixed orthogonal arrays with certain minimum Hamming distance, irredundant mixed orthogonal arrays and $k$-uniform states, we present two constructions of  $2$-uniform states in heterogeneous systems. We also construct a family of $3$-uniform states in  heterogeneous systems, which solves a question  posed in  [D. Goyeneche \emph{et al.}, \href{https://journals.aps.org/pra/abstract/10.1103/PhysRevA.94.012346}{Phys. Rev. A \textbf{94}, 012346 (2016)}]. We also show two methods of generating $(k-1)$-uniform states from $k$-uniform states. Some new results on the existence and nonexistence of absolutely maximally entangled states  are provided. For the applications, we present an orthogonal basis consisting of  $k$-uniform states with minimum support.  Moreover, we show that some $k$-uniform bases can not be distinguished by local operations and classical communications, and this shows quantum nonlocality with entanglement.

\end{abstract}

\maketitle


\section{Introduction}
\label{sec:int}
    Multipartite entanglement  plays a central role in quantum key distribution \cite{ekert1991quantum,gisin2002quantum,bennett1992quantum}, quantum
    teleportation \cite{bennett1993teleporting,bouwmeester1997experimental}  and quantum error correcting codes (QECCs) \cite{scott2004multipartite}. However, characterizing entanglement  in an arbitrary multipartite system can be challenging \cite{horodecki2009quantum}. Recently, a striking class of pure states called $k$-uniform states has attracted much attention. These states have the property that all of the reductions to $k$ parties are maximally mixed. Let $\bbC^d$ be the  $d$-dimensional Hilbert space. A homogeneous system associates with the Hilbert space $(\bbC^d)^{\otimes N}$. Suppose $\ket{\psi}$ is a $k$-uniform state in $(\bbC^d)^{\otimes N}$, then $k\leq \fl{N}{2}$ due to the Schmidt decomposition. Specially, a $\fl{N}{2}$-uniform state is called an absolutely maximally entangled (AME) state, and it is  maximally entangled across any bipartition. AME states can be used  for threshold quantum secret sharing schemes, for parallel and open-destination teleportation protocols \cite{helwig2012absolute,helwig2013absolutely}. AME states can also be used to design holographic
    quantum codes~\cite{pastawski2015holographic}. Further, a $k$-uniform state in  $(\bbC^d)^{\otimes N}$  corresponds to a pure QECC in $(\bbC^d)^{\otimes N}$ of distance $k+1$ \cite{scott2004multipartite}. Due to the various applications of $k$-uniform states, it is important to study them theoretically.

    In a more general case, a heterogeneous system associates with the Hilbert space $\bbC^{d_1}\otimes\bbC^{d_2}\otimes\cdots\otimes\bbC^{d_N}$.  In \cite{goyeneche2016multipartite}, the authors first studied $k$-uniform states in heterogeneous systems, and constructed several $1$-uniform and $2$-uniform states in heterogeneous systems  from  irredundant mixed orthogonal arrays ($\irm$s).  But we still do not know which mixed orthogonal array is irredundant. In \cite{bryan2018existence}, the authors gave a sufficient and necessary condition for the existence of $1$-uniform states in $\bbC^{d_1}\otimes\bbC^{d_2}\otimes\cdots\otimes\bbC^{d_N}$. However, there are few constructions of $k$-uniform states in heterogeneous systems when $k\geq 2$. Specially,  it was wondered in \cite{goyeneche2016multipartite} whether there exist $3$-uniform states in heterogeneous systems. We shall give a positive answer to this question and construct more $k$-uniform states in  heterogeneous systems. This is the first  motivation of this work.  AME states in tripartite heterogeneous systems were investigated in \cite{shen2020absolutely}. Recently, the authors in \cite{gu2019quantum} designed the quantum setups to produce some tripartite
   AME states in heterogeneous systems. It is reasonable to believe that AME states in heterogeneous systems can be experimentally realized in the near future. Thus, it is meaningful to study  AME states in  heterogeneous systems. This is the second motivation of this work.   Table~\ref{Table:k-unfiormresults} shows our main results  .

\begin{table}
	\renewcommand\arraystretch{1.7}	
	\caption{Existence of $k$-uniform states. Note that ``-'' means  unclear.}\label{Table:k-unfiormresults}
	\centering
	\renewcommand\tabcolsep{5pt}
	\begin{tabular}{ccccc}
		\midrule[1.1pt]
		$(\bbC^d)^{\otimes N}$  &Existence  &Nonexistence &Unknown   &References  \\
		\hline
		$1$-uniform  &$d\geq 2, N\geq 2$ &no  & no &\cite{goyeneche2014genuinely}  \\
		\hline
		$2$-uniform  &$d\geq 2,N\geq 4$ except $d=2,6$, $N=4$ &$d=2, N=4$ & $d=6$, $N=4$ &\cite{goyeneche2014genuinely,scott2004multipartite,li2019k,pang2019two,rains1999nonbinary,highu}  \\
		\hline
		\multirow{2}*{$3$-uniform} &\multirow{2}*{$d\geq 2$, $N\geq 6$,}  &\multirow{2}*{$d=2, N=7$} &\multirow{2}*{$d\geq 6$,  $d=2 \pmod 4$,}  &\multirow{2}*{\cite{li2019k,rains1999nonbinary,huber2017absolutely,helwig2013absolutely,raissi2019constructing,grassl2015quantum,pang2019two}}\\
		&except $d=2 \pmod 4$, $N=7$& &$N=7$ &\\
		
		\midrule[1.1pt]
		
		\multirow{2}*{$(\bbC^d)^{\otimes N}\otimes (\bbC^2)^{\otimes t}$}& \multirow{2}*{Existence}  &\multirow{2}*{Nonexistence} &\multirow{2}*{Unknown}   &\multirow{2}*{References}  \\ [5pt]
		\hline
		\multirow{4}*{$2$-uniform}
		& $d\geq 2$, $t=1$, $N\geq 5$&- &- &Theorem~\ref{Thm:dddd2} \\
		&\multirow{2}*{$d\geq 2$, $t=2$,}   &\multirow{2}*{-}  &\multirow{2}*{-}  &\multirow{2}*{Theorem~\ref{Thm:dddd2}}\\
		&$N\geq 7$,$N\neq 4d+2,4d+3$& & &\\

		&Table~\ref{Table:ct1ct2} &- &- &-\\
		\hline
		3-uniform&Proposition~\ref{pro:3uni} &- &- &-\\
		\hline
		AME states &- &Table~\ref{Table:nonAME} &- &-\\
		\midrule[1.1pt]
	\end{tabular}
\end{table}

   A set of orthogonal states is locally indistinguishable, if it is not possible to distinguish the states by any sequence of local operations and classical communications (LOCC).  Local indistinguishability can be used for data
   hiding \cite{terhal2001hiding,divincenzo2002quantum,eggeling2002hiding,Matthews2009Distinguishability} and quantum secret sharing \cite{Markham2008Graph}. Bennett \emph{et al.} first constructed an orthogonal basis  consisting of product states in $\bbC^3\otimes\bbC^3$ which is locally indistinguishable, and showed the phenomenon of quantum nonlocality without entanglement \cite{bennett1999quantum}. Recently, Halder \emph{et al.} proposed the concept of locally irreducible set, and showed the  $N$-qubit $\ghz$ bases are locally irreducible \cite{halder2019strong}. A
   set of orthogonal  states is locally irreducible if it is not possible to eliminate one or more  states
   from the set by nontrivial orthogonality-preserving local measurements.
   Local irreducibility sufficiently ensures
   local indistinguishability, but the converse is not ture. Since there are few results about the  local irreducibility of other entangled bases, it is meaningful to investigate the local irreducibility of  $k$-uniform bases. This is the third motivation of this work.

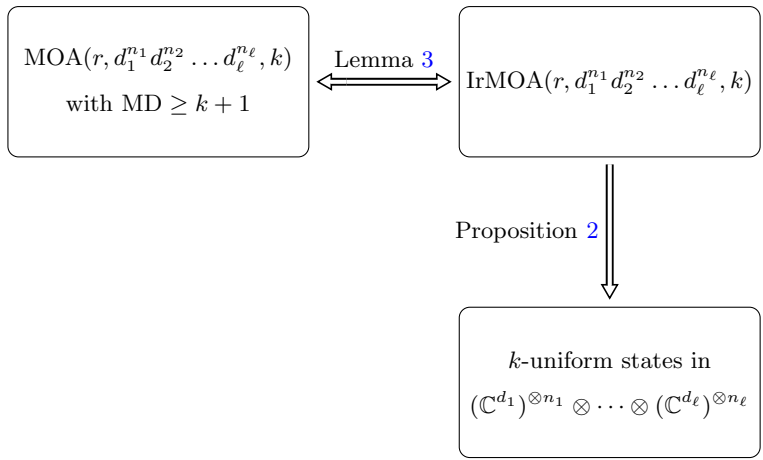
\begin{figure}[t]
	\begin{tikzpicture}
	\draw[rounded corners] (0,0) rectangle (4,2);
	\draw (2,1.3) node []{$\moa(r,d_1^{n_1}d_2^{n_2}\ldots d_{\ell}^{n_\ell},k)$};
	\draw (2,0.7) node []{with $\md\geq k+1$};
	
	\draw[rounded corners] (6,0) rectangle (10,2);
	\draw (8,1) node []{$\irm(r,d_1^{n_1}d_2^{n_2}\ldots d_{\ell}^{n_\ell},k)$};

	\draw[rounded corners] (6,-4) rectangle (10,-2);
	\draw (8,-2.7) node []{$k$-uniform states in};
	\draw (8,-3.3) node []{$(\bbC^{d_1})^{\otimes n_1}\otimes \cdots\otimes(\bbC^{d_\ell})^{\otimes n_\ell}$};

	\draw [vecArrow](4.5,1)--(5.9,1);
	\draw [vecArrow](4.5,1)--(4.1,1);
	\draw (5,1.3) node []{Lemma \ref{Lemma:mdmoa}};
	\draw [vecArrow](8,-0.1)--(8,-1.9);
	\draw (6.9,-1) node []{Proposition \ref{pp:moaandkuniform}};
	\end{tikzpicture}
	\caption{The main method  of constructing $k$-uniform states in heterogeneous systems in this paper. }\label{Fig:moairmoa}
\end{figure}

  In this paper, we investigate $k$-uniform states in heterogeneous systems for $k=2,3$. Similar to homogeneous $k$-uniform states,   heterogeneous $k$-uniform states are related to QECCs
  over mixed alphabets \cite{goyeneche2016multipartite,wang2013quantum}.  In Proposition~\ref{pp:moaandkuniform}, we first restate the connection between $\irm$s and $k$-uniform states in heterogeneous systems. Second we give an efficient way to check whether an $\moa$ is irredundant in Lemma~\ref{Lemma:mdmoa}.  Figure~\ref{Fig:moairmoa} shows our main method  of constructing $k$-uniform states in heterogeneous systems. We establish  Constructions~\ref{ct:replacement} and ~\ref{ct:diffenencesch}, and we use these two constructions to construct $2$-uniform states in heterogeneous systems in Theorem~\ref{Thm:dddd2} and Table~\ref{Table:ct1ct2}. Figure~\ref{Fig:expan} shows our Constructions~\ref{ct:replacement}. Next, we propose Construction~\ref{ct:d3}, and we use it to construct a family of $3$-uniform states in  heterogeneous systems in Proposition~\ref{pro:3uni}. Moreover, we show two methods of generating $(k-1)$-uniform states from $k$-uniform states in heterogeneous systems in Propositions~\ref{pro:projective} and \ref{Pro:k+1-k}.  We show the existence of AME states in Corollary~\ref{cor:AME}, and the nonexistence  of AME states in Lemma~\ref{lem:ame3128} and Table~\ref{Table:nonAME}.   Finally, we indicate some applications of $k$-uniform states in heterogeneous systems. We  present an orthogonal basis consisting of  $k$-uniform states with minimum support in  heterogeneous system in Proposition~\ref{Pro:k-basis}. We also show some $k$-uniform bases are locally irreducible in Proposition~\ref{pro:local} and Corollary~\ref{cor:ghzbasis}.  Figure~\ref{Fig:mini} shows  all of our applications.

  It is known that  $k$-uniform states in homogeneous systems can be constructed from orthogonal arrays, Latin squares, symmetric matrices, graph states, quantum error correcting codes and classical error correcting codes   \cite{goyeneche2014genuinely,feng2017multipartite,goyeneche2015absolutely,scott2004multipartite,goyeneche2018entanglement,helwig2013absolutelygraph}. Table~\ref{Table:k-unfiormresults} shows the existence $1,2,3$-uniform states in homogeneous systemsA.  Multiqubit AME states only exist for $2,3,5,6$-qubits \cite{scott2004multipartite,rains1999quantum,huber2017absolutely}. There are still some unknown AME states for local dimension $d\geq 3$ \cite{AMEtable}. Specially, the  existence of AME states in $\bbC^6\otimes\bbC^6\otimes\bbC^6\otimes\bbC^6$ remains open \cite{horodecki2020five}. Multipartite entanglement in heterogeneous systems has been studied in  \cite{huber2013structure,sun2015classification,chen2006classification,chen2006range,miyake2004multipartite,yu2008genuine}. In \cite{krenn2016automated}, the authors gave an algorithm to design quantum experiments in heterogeneous systems. The capacity of quantum channels  and efficiency
  of quantum gates can be increased, if we take into account each
  subsystem consisting of more than two levels  \cite{ralph2007efficient,fujiwara2003exceeding}. However, the study
  of entanglement for heterogeneous systems is  more
  complicated  than that for homogeneous systems, due to the lack of useful
  mathematical tools \cite{goyeneche2016multipartite}.

  The remainder of this paper is organized as follows. In Sec. \ref{sec:preli}, we introduce the preliminary knowledge and facts. In Sec. \ref{sec:con}, we construct $k$-uniform states in heterogeneous systems.  AME states in homogeneous systems are investigated in Sec. \ref{sec:AMEkuniform}.  In Section~\ref{sec:twoapp}, we indicate some applications. We conclude in Sec. \ref{sec:conclusionpart}. In addition, we give a proof of Construction~\ref{ct:diffenencesch} in Appendix~\ref{Appendix:A},  a proof of Theorem~\ref{Thm:dddd2} in Appendix~\ref{Appendix:D}, a proof of Proposition~\ref{pro:projective} in Appendix~\ref{Appendix:E},a proof of Proposition~\ref{Pro:k-basis} in Appendix~\ref{Appendix:B}, and a proof of Proposition~\ref{pro:local} in Appendix~\ref{Appendix:C}.

\section{Preliminary}
\label{sec:preli}
In this section we introduce the preliminary knowledge
and facts used in this paper. Let
$(\bbC^d)^{\otimes N}$ denote $\bbC^d\otimes\bbC^d\otimes\cdots\otimes\bbC^d$, where $d$ repeats $N$ times. For convenience, we assume that $d_1\geq d_2\geq \cdots\geq d_N$ in $\bbC^{d_1}\otimes\bbC^{d_2}\otimes\cdots\otimes\bbC^{d_N}$.
A \emph{$k$-uniform state} $\ket{\psi}$ in $\bbC^{d_1}\otimes\bbC^{d_2}\otimes\cdots\otimes\bbC^{d_N}$ has the property that all reductions to $k$ parties are maximally mixed. That is,	for any subset $\{i_1,i_2,\ldots,i_k\}\subset \{1,2,\ldots,N\}$, we have
	\begin{equation}
	\rho_{\{i_1,i_2,\ldots,i_k\}}=\tr_{\{i_1,i_2,\ldots,i_k\}^c}\ketbra{\psi}{\psi}=\frac{1}{d_{i_1}d_{i_2}\cdots d_{i_k}}I_{d_{i_1}d_{i_2}\cdots d_{i_k}},
	\end{equation}
	where
 $\{i_1,i_2,\ldots,i_k\}^c=\{1,2,\ldots,N\}/\{i_1,i_2,\ldots,i_k\}$, and $\tr_{\{i_1,i_2,\ldots,i_k\}^c}$ is the partial trace operation. Due to the Schmidt decomposition of bipartite pure state, we obtain that $d_1d_2\cdots d_k\leq d_{k+1}d_{k+2}\cdots d_N$. Thus $k$ satisfies $k\leq \fl{N}{2}$.  If $k=\fl{N}{2}$ then $\ket{\ps}$ is also called an absolutely maximally entangled (AME) state.

Next, \emph{orthogonal arrays} and \emph{mixed orthogonal arrays} are  essential in statistics
and have wide applications in computer science and cryptography. They are related to  finite fields, finite geometry and classical error-correcting codes \cite{hedayat1999orthogonal}. Formally, a mixed orthogonal array $\moa(r,d_1^{n_1}d_2^{n_2}\ldots d_\ell^{n_l},k)$ is an array of $r$ rows and $N$ columns, where $N=\sum_{i=1}^\ell n_i$,  the first $n_1$ columns have symbols from $\{0,1,\ldots,d_1-1\}$, the next $n_2$
columns have symbols from $\{0,1,\ldots,d_2-1\}$, and so on, with the property that
in any $r\times k$ subarray every possible $k$-tuple occurs the same number of times
as a row. Here each $d_i$ is called a level, and $d_i^{n_i}$ means level $d_i$ repeats $n_i$ times. If $\ell=1$ and $n_1=N$, then a mixed orthogonal array is reduced to an orthogonal array, denoted by $\oa(r,d^N,k)$. Without loss of generality, we assume $d_1> d_2> \cdots> d_l$. A mixed orthogonal array is \emph{simple} if the rows are all distinct.
 The following are two examples of simple $\moa$s.
\begin{example}\label{Ex:moa12}
$\begin{pmatrix}
 0 &0  &0  &0  &0\\
 0 &1  &1  &1  &1\\
 1 &0  &0  &1  &1\\
 1 &1  &1  &0  &0\\
 2 &0  &1  &0  &1\\
 2 &1  &0  &1  &0\\
 3 &0  &1  &1  &0\\
 3 &1  &0  &0  &1\\
\end{pmatrix} \ \text{is an} \ \moa(8,4^12^4,2); \ \ \ \ \ \ \ \ \ \ \ \
\begin{pmatrix}
0&0&0&0&0\\
0&1&0&1&0\\
0&0&1&0&1\\
0&1&1&1&1\\
1&0&1&1&0\\
1&1&1&0&0\\
1&0&0&0&1\\
1&1&0&1&1\\
2&0&1&1&0\\
2&1&0&0&0\\
2&0&0&1&1\\
2&1&1&0&1\\
\end{pmatrix} \ \text{is an} \ \moa(12,3^12^4,2)$.
\end{example}
Using the $\moa(8,4^12^4,2)$ in the above example, we can construct a $2$-uniform state in  $\bbC^4\otimes (\bbC^2)^{\otimes4}$,
\begin{equation}\label{eq:42222}
\begin{split}
\ket{\psi}=&\frac{1}{2\sqrt{2}}(\ket{00000}+\ket{01111}+\ket{10011}+\ket{11100}\\
&+\ket{20101}+\ket{21010}+\ket{30110}+\ket{31001}).
\end{split}
\end{equation}
Not all mixed orthogonal arrays can be used to construct $k$-uniform states in heterogeneous systems. The $\moa(12,3^12^4,2)$ in Example \ref{Ex:moa12} implies the state $\ket{\phi}=\frac{1}{2\sqrt{3}}(\ket{00000}+\ket{01010}+\ket{00101}+\ket{01111}+\ket{10110}+\ket{11100}+\ket{10001}+\ket{11011}+\ket{20110}+\ket{21000}+\ket{20011}+\ket{
21101})\in \bbC^3\otimes (\bbC^2)^{\otimes4}$. However, $\ket{\phi}$ is not a $2$-uniform state, because the reduced density operator of the first two parties of $\ket{\phi}$ is no longer diagonal. To characterize the relation between mixed orthogonal arrays and $k$-uniform states, we introduce the \emph{irredundant} mixed orthogonal arrays \cite{goyeneche2016multipartite}.
An $\moa(r,d_1^{n_1}d_2^{n_2}\ldots d_\ell^{n_\ell},k)$ with $N=\sum_{i=1}^\ell n_i$ columns is called irredundant,  if when removing from the array any $k$ columns, all
	remaining $r$ rows containing $N-k$ symbols each, are all
	different. Denote $\irm(r,d_1^{n_1}d_2^{n_2}\ldots d_\ell^{n_\ell},k)$ and $\iro(r,d^N,k)$  as the irredundant $\moa(r,d_1^{n_1}d_2^{n_2}\ldots d_\ell^{n_\ell},k)$ and irredundant $\oa(r,d^N,k)$, respectively. In the following, we show that irredundant mixed orthogonal arrays can be used to construct $k$-uniform states in heterogeneous systems. The proof is given in \cite{goyeneche2016multipartite}.
\begin{proposition}
\label{pp:moaandkuniform}
	If array $(m_{i,j})_{1\leq i\leq r;1\leq j\leq N}$ is an $\irm(r,d_1^{n_1}d_2^{n_2}\ldots d_\ell^{n_\ell},k)$, then
	$
	\ket{\psi}=\frac{1}{\sqrt r}\sum_{i=1}^r\ket{m_{i,1}m_{i,2}\ldots m_{i,N}}
$ is a $k$-uniform state in $(\bbC^{d_1})^{\otimes n_1}\otimes (\bbC^{d_2})^{\otimes n_2}\otimes\cdots\otimes(\bbC^{d_\ell})^{\otimes n_\ell}$.
\end{proposition}

In Example~\ref{Ex:moa12}, one can verify that the $\moa(8,4^12^4,2)$ is irredundant, while the $\moa(12,3^12^4,2)$ is not. By Proposition \ref{pp:moaandkuniform} we can  construct a $2$-uniform state in  $\bbC^{4}\otimes(\bbC^{2})^{\otimes 4}$. In the next section, we give some constructions of $k$-uniform states in heterogeneous systems by irredundant mixed orthogonal arrays.

\section{Constructions of $k$-uniform states in heterogeneous systems}\label{sec:con}
In this section, we introduce the connection between mixed orthogonal arrays with certain minimum Hamming distance and irredundant mixed orthogonal arrays in Lemmas~\ref{Lemma:mdmoa} and \ref{Lemma:deletec}. Based on this connection, we give Constructions~\ref{ct:replacement} and \ref{ct:diffenencesch}, and we use these two constructions to construct $2$-uniform states in heterogeneous systems in Theorem~\ref{Thm:dddd2} and Table~\ref{Table:ct1ct2}. We also give Construction~\ref{ct:d3}, and we use it to construct $3$-uniform states in heterogeneous systems in Proposition~\ref{pro:3uni}. We also give two methods of generating $(k-1)$-uniform states from $k$-uniform states in Propositions~\ref{pro:projective} and \ref{Pro:k+1-k}.

\subsection{The minimum Hamming distance of $\moa$s}

It is not easy to check whether an $\moa$ is an $\irm$ by definition when it has many columns. We introduce an efficient way to check whether
an  $\moa$ is irredundant. This  method was first used to check whether
an  $\oa$ is irredundant \cite{pang2019two}. First of all, we introduce the \emph{Hamming distance} of two vectors, which is from the literature of coding theory. For an $r\times N$ matrix $M=\{\bm_1,\bm_2,\ldots,\bm_r\}$, the Hamming distance between the row vectors $\bm_i$ and  $\bm_j$, denoted by $\hd(\bm_i,\bm_j)$, is the number of column positions in which symbols are different. We call $\min\limits_{1\leq i<j\leq r}\hd(\bm_i,\bm_j)$ the minimum Hamming distance of $M$, denoted by $\md(M)$. Now we give the connection  between $\moa$s with certain $\md$ and $\irm$s.
\begin{lemma}\label{Lemma:mdmoa}
	If $M$ is an $\moa(r,d_1^{n_1}d_2^{n_2}\ldots d_\ell^{n_\ell},k)$, then $M$ is irredundant if and only if $\md(M)\geq k+1$.
\end{lemma}
\begin{proof}
The sufficiency is obvious. We only need to consider the necessity. Let $N=n_1+n_2+\cdots+n_\ell$. Assume   $\hd(\bm_i,\bm_j)$=$\md(M)=b\leq k$, where $\bm_i=(m_{i,s})_{1\leq s\leq N},\bm_j=(m_{j,s})_{1\leq s\leq N}$ are two row vectors of $M$, then there exists a set $\{s_1,s_2,\ldots,s_b\}\subset \{1,2,\ldots,N\}$ such that $m_{i,s_t}\neq m_{j,s_t}$ for any $1\leq t\leq b$. It means that for any  $s\in \{s_1,s_2,\ldots,s_b\}^c$, $m_{i,s}= m_{j,s}$. Since $N-b\geq N-k$, it contradicts to the definition of $\irm$s.
\end{proof}
\vspace{0.5cm}

Specially, an $\moa$ is simple if and only if $\md\geq 1$.
 Figure~\ref{Fig:moairmoa} shows our main method of constructing $k$-uniform states in heterogeneous systems.
The $\md$s of the $\moa(8,4^12^4,2)$ and the $\moa(12,3^12^4,2)$ given by Example~\ref{Ex:moa12} are $3$ and $1$ respectively.  Thus the $\moa(8,4^12^4,2)$ is irredundant, while the $\moa(12,3^12^4,2)$ is not by Lemma~\ref{Lemma:mdmoa}.

If we remove some columns of $\moa(r,d_1^{n_1}d_2^{n_2}\ldots d_\ell^{n_\ell},k)$, it is still an $\moa(r,d_1^{n_1'}d_2^{n_2'}\ldots d_\ell^{n_\ell'},k)$. In ~\cite{goyeneche2016multipartite}, the authors proposed a concept called \emph{endurance of $k$ uniformity}, which is the maximum number of columns of an $\irm$ with strength $k$ that can be removed so that the resultant $\moa$ preserves
both irredundancy and strength $k$.   By Lemma~\ref{Lemma:mdmoa}, we can easily estimate the endurance of $k$ uniformity, which must be at least $\md-(k+1)$. Thus we have the following lemma.

\begin{lemma}\label{Lemma:deletec}
Suppose $M$ is an $\irm(r,d_1^{n_1}d_2^{n_2}\ldots d_\ell^{n_\ell},k)$ with $\md=b\geq k+1$. If we remove any $c\leq b-(k+1)$ columns of $M$, then it becomes an $\irm(r,d_1^{n_1'}d_2^{n_2'}\ldots d_\ell^{n_\ell'},k)$, where $n_i'\leq n_i$, $1\leq i\leq \ell$ and  $(n_1+n_2+\cdots+n_\ell)-(n_1'+n_2'+\cdots+n_\ell')=c$.
\end{lemma}

Now, we give an example.
\begin{example}\label{Ex:moa18}
$$\left(\begin{array}{cccccccccccccccccc}
0  &0   &0   &0    &0   &0   &1  &1  &1  &1   &1   &1   &2  &2   &2  &2  &2   &2\\
0  &1   &2   &1    &2   &0   &1  &2  &0  &2   &0   &1   &2  &0   &1  &0  &1   &2\\
0  &2   &1   &1    &0   &2   &1  &0  &2  &2   &1   &0   &2  &1   &0  &0  &2   &1\\
0  &2   &2   &0    &1   &1   &1  &0  &0  &1   &2   &2   &2  &1   &1  &2  &0   &0\\
0  &0   &1   &2    &2   &1   &1  &1  &2  &0   &0   &2   &2  &2   &0  &1  &1   &0\\
0  &1   &0   &2    &1   &2   &1  &2  &1  &0   &2   &0   &2  &0   &2  &1  &0   &1\\
0  &1   &2   &0    &1   &2   &0  &1  &2  &0   &1   &2   &0  &1   &2  &0  &1   &2\\
0  &0   &0   &1    &1   &1   &0  &0  &0  &1   &1   &1   &0  &0   &0  &1  &1   &1\\
\end{array}\right)^{T}\ \text{is an} \ \moa(18,3^72^1,2).$$
It has $\md=5$, so we can obtain an $\irm(18,3^62^1,2)$, an $\iro(18,3^7,2)$, an $\irm(18,3^52^1,2)$ and an $\iro(18,3^6,2)$ by Lemma~\ref{Lemma:deletec}.
\end{example}

\subsection{Constructions of 2-uniform states in heterogeneous systems}
\label{ssec:construction2}
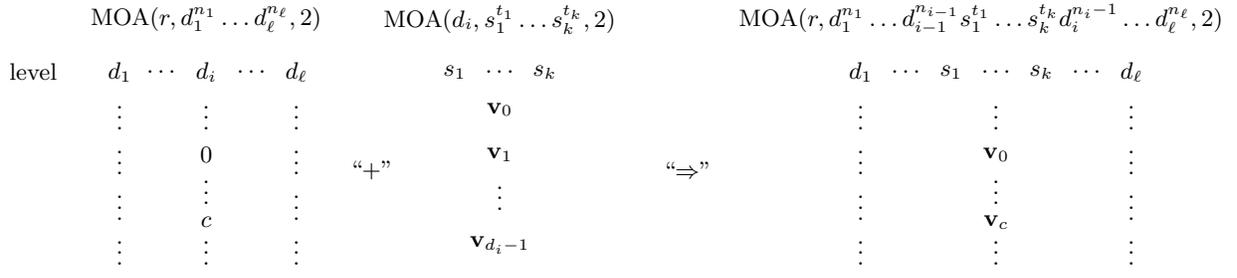
\begin{figure}
	\begin{tikzpicture}
	\draw (-0.9,0) node []{$\moa(r, d_1^{n_1}\ldots d_\ell^{n_\ell},2)$};
    \draw (-3.2,-0.7) node []{level};
    \draw (-2.05,-0.7) node []{$d_1$};
    \draw (-2.05,-1.2) node []{$\vdots$};
    \draw (-2.05,-1.8) node []{$\vdots$};
    \draw (-2.05,-2.4) node []{$\vdots$};
    \draw (-2.05,-3) node []{$\vdots$};
    \draw (-1.5,-0.7) node []{$\ldots$};
    \draw (-0.9,-0.7) node []{$d_i$};
    \draw (-0.9,-1.2) node []{$\vdots$};
    \draw (-0.9,-1.8) node []{$0$};
    \draw (-0.9,-2.2) node []{$\vdots$};
    \draw (-0.9,-2.7) node []{$c$};
    \draw (-0.9,-3) node []{$\vdots$};
    \draw (-0.3,-0.7) node []{$\ldots$};
    \draw (0.3,-0.7) node []{$d_\ell$};
    \draw (0.3,-1.2) node []{$\vdots$};
    \draw (0.3,-1.8) node []{$\vdots$};
    \draw (0.3,-2.4) node []{$\vdots$};
    \draw (0.3,-3) node []{$\vdots$};

    \draw (1.3,-2) node []{``+''};

    \draw (3,0) node []{$\moa(d_i,s_1^{t_1}\ldots s_k^{t_k},2)$};
    \draw (2.4,-0.7) node []{$s_1$};
    \draw (3,-0.7) node []{$\ldots$};
    \draw (3,-1.2) node []{$\bv_0$};
    \draw (3,-1.8) node []{$\bv_1$};
    \draw (3,-2.27) node []{$\vdots$};
    \draw (3,-3) node []{$\bv_{d_i-1}$};
    \draw (3.6,-0.7) node []{$s_k$};

    \draw (5.5,-2) node []{``$\Rightarrow$''};

 \draw (9.4,0) node []{$\moa(r,d_1^{n_1}\ldots d_{i-1}^{n_{i-1}}s_1^{t_1}\ldots  s_k^{t_k}d_i^{n_i-1} \ldots d_\ell^{n_\ell},2)$};

\draw (7.8,-0.7) node []{$d_1$};
\draw (7.8,-1.2) node []{$\vdots$};
\draw (7.8,-1.8) node []{$\vdots$};
\draw (7.8,-2.4) node []{$\vdots$};
\draw (7.8,-3) node []{$\vdots$};
\draw (8.4,-0.7) node []{$\ldots$};

\draw (9,-0.7) node []{$s_1$};
\draw (9.6,-0.7) node []{$\ldots$};
\draw (9.6,-1.2) node []{$\vdots$};
\draw (9.6,-1.8) node []{$\bv_0$};
\draw (9.6,-2.2) node []{$\vdots$};
\draw (9.6,-2.7) node []{$\bv_{c}$};
\draw (9.6,-3) node []{$\vdots$};
\draw (10.2,-0.7) node []{$s_k$};

\draw (10.8,-0.7) node []{$\ldots$};
\draw (11.4,-0.7) node []{$d_\ell$};
\draw (11.4,-1.2) node []{$\vdots$};
\draw (11.4,-1.8) node []{$\vdots$};
\draw (11.4,-2.4) node []{$\vdots$};
\draw (11.4,-3) node []{$\vdots$};
	\end{tikzpicture}
	\caption{Expansive replacement. We replace one column of level $d_i$ of $\moa(r, d_1^{n_1}\ldots d_\ell^{n_\ell},2)$ with a simple $\moa(d_i,s_1^{t_1}\ldots s_k^{t_k},2)$. It means that we replace  the symbol $c$ in the column of level $d_i$ with vector $\bv_c$ for $c=0,1,\ldots,d_i-1$. Then we obtain an $\moa(r,d_1^{n_1}\ldots d_{i-1}^{n_{i-1}}s_1^{t_1}\ldots  s_k^{t_k}d_i^{n_i-1} \ldots d_\ell^{n_\ell},2)$.  }\label{Fig:expan}
\end{figure}

In this subsection, we introduce some methods of constructing $\moa(r,d_1^{n_1}d_2^{n_2}\ldots d_\ell^{n_\ell},2)$ with certain $\md$. The first method is the so-called expansive replacement \cite[Chapter 9]{hedayat1999orthogonal}. See the following construction and Figure~\ref{Fig:expan}.

\begin{construction}
\label{ct:replacement}
Suppose $A_1$ is an $\moa(r, d_1^{n_1}d_2^{n_2}\ldots d_\ell^{n_\ell},2)$ with $\md(A_1)= b$, and $A_2$ is a simple $\moa(d_i,s_1^{t_1}s_2^{t_2}\ldots s_k^{t_k},2)$.
Suppose  $\bf{c}$ is a column vector of $A_1$ with level $d_i$. Namely $\bf{c}$ is the $(n_1+n_2+\ldots+n_{i-1}+k)$-th column of $A_1$ for some $1\leq k\leq n_i$. Assume $A_2$ has row vectors $\{\bv_0,\bv_1,\ldots,\bv_{d_i-1}\}$. If we replace the symbol $c$ of $\bf{c}$ with $\bv_c$ for $c=0,1,...,d_i-1$,
then we obtain  $A_1'$ which is an $\moa(r,d_1^{n_1}d_2^{n_2}\ldots d_{i-1}^{n_{i-1}}s_1^{t_1}s_2^{t_2}\ldots  s_k^{t_k}d_i^{n_i-1}d_{i+1}^{n_{i+1}} \ldots d_\ell^{n_\ell},2)$ with $\md(A_1')\geq b$. Thus, if $A_1$ is irredundant, then $A_1'$ must be irredundant.
\end{construction}
\begin{proof}
	We only need to discuss the $\md$ of $A_1'$. Assume  $\hd(\bm_i,\bm_j)=b$, where $\bm_i=(m_{i,s}),\bm_j=(m_{j,s})$ are two row vectors of $A_1$. Denote $\bm_i'$ (resp. $\bm_j'$) is the replaced row  of $\bm_{i}$ (resp. $\bm_{j}$) in $A_1'$. If the elements of $m_{i,s}$ and $m_{j,s}$ in level $d_i$ (replaced column) are the same, then after replacement,  $\hd(\bm_i',\bm_j')=b$.  If the elements of $m_{i,s}$ and $m_{j,s}$ in level $d_i$ (replaced column) are different, then $\hd(\bm_i',\bm_j')\geq b$, since $A_2$ is simple. Thus, we have $\md(A_1')\geq b$.
\end{proof}
\vspace{0.5cm}
	
	For example, there is an $\moa(36,12^13^{12},2)$ with $\md=9$ and a simple
	$\moa(12,3^12^4,2)$ \cite{Orthogaonaltable}, then we obtain an $\moa(36,3^{13}2^4,2)$ with $\md\geq 9$ by Construction \ref{ct:replacement}.
	
    There is a \emph{splitting} method for $\moa$s in \cite{brouwer2006orthogonal}. Given an $\moa(r,(d_1d_2)^1d_3^1\ldots d_N^1,t)$ with $\md=b$, where $d_1d_2,d_3,\ldots,d_N$ are unordered, we can obtain an $\moa(r,d_1^1d_2^1d_3^1\ldots d_N^1,t)$ with $\md\geq b$, by replacing the symbols in level $d_1d_2$ with those pairs of a trivial $\moa(d_1d_2,d_1^1d_2^1,2)$. The discussion of $\md$ is the same as the proof of Construction~\ref{ct:replacement}.
	For example, there is an $\moa(18,3^66^1,2)$ with $\md=5$  \cite{Orthogaonaltable}, then we obtain an $\moa(18,3^72^1,2)$ with $\md\geq 5$. It is worth noting that there exists also a splitting method for $k$-uniform states \cite{goyeneche2016multipartite}. Given  a $k$-uniform state in $\bbC^{d_1d_2}\otimes\bbC^{d_3}\otimes\cdots\otimes\bbC^{d_N}$,  we can obtain  a $k$-uniform state in $\bbC^{d_1}\otimes\bbC^{d_2}\otimes\bbC^{d_3}\otimes\cdots\otimes\bbC^{d_n}$, by replacing the states in $\bbC^{d_1d_2}$ with those in $\bbC^{d_1}\otimes\bbC^{d_2}$. By this reason, the authors in \cite{goyeneche2016multipartite} mainly focused on \emph{genuinely heterogeneous systems}, i.e., systems composed of subsystems with coprime levels (e.g., qubits-qutrits or qutrits-ququints). We emphasize that the converse of the splitting method is not true. For example, we cannot obtain a $1$-unfiorm state in $\bbC^4\otimes\bbC^2\otimes\bbC^2$ from a $\ket{\ghz}=\frac{1}{\sqrt{2}}(\ket{0000}+\ket{1111})$ in $(\bbC^2)^{\otimes 4}$. The converse of the splitting method can be true if we add a condition. We will discuss this probem in Proposition~\ref{Pro:k+1-k}.  Thus, we focus on the  heterogeneous systems no matter whether they are genuinely heterogeneous systems.
	
Next, we give the second construction of $\moa(r,d_1^{n_1}d_2^{n_2}\ldots d_\ell^{n_\ell},2)$ with certain $\md$ by \emph{difference scheme}. Let $S$ be an additive group with $d$ different symbols. An $s\times N$ matrix $D$ with entries from $S$ is called a difference scheme if it has the property that the difference between every pair of columns contains the $d$ symbols equally often \cite[Chapter 6]{hedayat1999orthogonal}. Denote it by $D(s,N,d)$.  We often choose $S=\bbZ_d$ or \emph{Galois field} GF$(d)$. A difference scheme $D(\lambda d,\lambda d,d)$ is also called a \emph{generalized Hadamard matrix} ($\ghm$). Since $S$ is an additive group, the transpose of $D(\lambda d,\lambda d,d)$ is still a $\ghm$.

For two matrices $A_1=(a_{i,j})_{1\leq i\leq r_1,1\leq j\leq N_1}$ and $A_2=(f_{i,j})_{1\leq i\leq r_2,1\leq j\leq N_2}$, denote $A_1\oplus A_2=(a_{i,j}+A_2)_{{1\leq i\leq r_1,1\leq j\leq N_1}}$,  where $a_{i,j}+A_2=(a_{i,j}+f_{i',j'})_{1\leq i'\leq r_2,1\leq j'\leq N_2}$. Thus $A_1\oplus A_2$ has $r_1r_2$ rows and $N_1N_2$ columns. Given an $\oa(r,d^{N},2)$, $N\geq 1$ and a $\ghm$ $D(\lambda d,\lambda d,d)$, we can obtain a new $\moa$ with certain $\md$ through ``$\oplus$''. Note that when $N=1$, we say $(
0,1,\ldots,d-1)^T$ is an $\oa(d,d^1,2)$.
	
\begin{construction}\label{ct:diffenencesch}
   Let $A_1$ be an $\oa(r,d^{N},2)$ with $\md(A_1)=b$, $N\geq 1$.  Assume $A_2$ is a $\ghm$ $D(\lambda d,\lambda d,d)$, then $A_1'=(
		\bm^T, A_1\oplus A_2
	)$ is an $\moa(r\lambda d,(\lambda d)^1d^{N\lambda d},2)$ with $\md(A_1')= \min\{\lambda(d-1)N+1, \lambda db\}$,  where the row vector $\bm=(m_j)_{1\leq j\leq r\lambda d}$ has $m_j=j-1 \pmod {\lambda d}$.
\end{construction}

See Appendix~\ref{Appendix:A} for the proof of Construction~\ref{ct:diffenencesch}. Now, we give some results of $2$-uniform states by Construction~\ref{ct:diffenencesch}.
\begin{theorem}\label{Thm:dddd2}
 \begin{enumerate}[(i)]
 	\item For any $d\geq 2$, there exists a $2$-uniform state in $(\bbC^d)^{\otimes N}\otimes\bbC^2\otimes\bbC^2$ for any $N\geq 7$ and $N\neq 4d+2,4d+3$.
 	\item For any $d\geq 2$, there exists a 2-uniform state in $(\bbC^d)^{\otimes N}\otimes\bbC^2$ for any  $N\geq 5$.
 \end{enumerate}	
\end{theorem}

See Appendix~\ref{Appendix:D} for the proof of Theorem~\ref{Thm:dddd2}. Other types of $\ghm$s can be found in \cite{colbourn2006crc,Differ}.  Let $A_2$ be a $D(s,N_1,d)$ in Construction~\ref{ct:diffenencesch}, then $A_1'=(
\bm^T,A_1\oplus A_2)$ is an $\moa(rs,s^1d^{NN_1},2)$, where vector $\bm=(m_j)_{1\leq j\leq rs}$ has $m_j=j-1$ $\pmod s$ \cite{hedayat1999orthogonal}. Although we cannot determine the $\md$ of $A_1'$ directly, we still can use it to construct $\irm$s. For example,  there exists a  difference scheme $D(15,9,3)$ $A_2$ \cite{Differ}. Let $A_1=(
0,1,2)^T$, then  $A_1'=(\bm^T,A_1\oplus A_2)$ is an $\moa(45,15^13^{9},2)$. We numerically determine $\md(A_1')=5$. Thus, we can construct a $2$-uniform state in $\bbC^5\otimes(\bbC^{3})^{\otimes{N}}$ for $8\leq N\leq 10$.

 Obviously, we can obtain more constructions of $2$-uniform states in heterogeneous systems by combining Constructions~\ref{ct:replacement} and \ref{ct:diffenencesch}.
\begin{example}\label{ex:repl}
	There exists a 2-uniform state in $(\bbC^3)^{\otimes N}\otimes (\bbC^2)^{\otimes t}$ if $t$ and $N$ satisfy one of the following two conditions:
	(a) $3\leq t \leq 4$, $N\geq 7$,
	(b) $5\leq t \leq 11$, $N\geq 6$. We prove the existence as follows.
	
	From the proof of Theorem~\ref{Thm:dddd2} (i), we have an $\moa(4d^{n+1},(4d^n)^1d^{4d^n},2)$  with $\md=4d^{n-1}(d-1)+1$ for $n\geq 1$, and an $\moa(4d^{n+1},(4d)^1d^{\frac{4d(d^n-1)}{d-1}},2)$ with $\md=4d^n-3$ for $n\geq 2$,  where $d$ is a prime power. We can obtain an  $\moa(4d^{n+1},(4d)^1d^{4d^n+n-1},2)$  with $\md\geq 4d^{n-1}(d-1)+1$ by splitting method.
	Thus when $d$ is a prime power, there exists an $\irm(r,(4d)^1d^N,2)$ $A_1$ for $N\geq 6$ and $N\neq 4d+1$, $4d+2$, by the same discussion as Theorem~\ref{Thm:dddd2} (i).
	
    Since there exists an $\moa(72,3^{24}24^1,2)$ with $\md=17$ \cite{Orthogaonaltable},  there exists an $\irm(72,(12)^13^N,2)$ for $11\leq N\leq 24$ by splitting method and Lemma~\ref{Lemma:deletec}. Let $d=3$ in $A_1$, then there exists an $\irm(r,(12)^13^N,2)$ for $N\geq 6$ and $N\neq 13,14$. Since $11\leq 13,14\leq 24$,   there exists an  $\irm(r,(12)^13^N,2)$ $A_1'$ for $N\geq 6$.
	 There exists a simple $\moa(12,3^12^t,2)$ $A_2$ for $3\leq t\leq 4$ (Example~\ref{Ex:moa12} gives a simple $\moa(12,3^12^4,2)$. It becomes a simple  $\moa(12,3^12^3,2)$ if we remove the last column of $\moa(12,3^12^4,2)$). We can obtain an $\irm(r,3^{N+1}2^t,2)$ for $3\leq t\leq 4$ and $N\geq 6$,  by combining $A_1'$ and $A_2$ through Construction~\ref{ct:replacement}.
	Moreover, there exists a simple $\oa(12,2^t,2)$ for $5 \leq t\leq 11$ \cite{Orthogaonaltable}. Thus, there exists an $\irm(r,3^{N}2^t,2)$ for  $5 \leq t\leq 11$ and $N\geq 6$.
\end{example}

\begin{table}
	\renewcommand\arraystretch{1.7}	
	\caption{$2$-uniform states in heterogeneous systems by combining Constructions~\ref{ct:replacement} and \ref{ct:diffenencesch}}
	\label{Table:ct1ct2}
	\centering
	\renewcommand\tabcolsep{6.0pt}
	\begin{tabular}{cccc}
		\hline
		$\irm(r,(4d)^1d^N,2)$ \ \ +&simple $\moa$s &$\Rightarrow$\ \ $\irm(r,d^N2^t,2)$ \\
		\hline
		\multirow{2}*{$d=3$, $N\geq 6$}& $\moa(12,3^12^t,2)$,  $3\leq t\leq 4$ &$d=3$, $3\leq t\leq 4$, $N\geq 7$\\
		&$\oa(12,2^t,2)$, $5\leq t\leq 11$ &$d=3$, $5\leq t\leq 11$, $N\geq 6$ \\	
		\hline
		\multirow{3}*{$d=5$, $N\geq 6$, $N\neq 21,22$}& \multirow{2}*{$\moa(20,5^12^t,2)$, $3\leq t\leq 8$} & $d=5$, $3\leq t \leq 4$, $N\geq 7$, $N\neq 22$, $23$ \\
		& \multirow{2}*{$\oa(20,2^t,2)$, $5\leq t\leq 19$} &$d=5$, $5\leq t \leq 8$, $N\geq 6$, $N\neq 22$\\
		& &$d=5$, $9\leq t \leq 19$, $N\geq 6$, $N\neq 21,22$\\	
		\hline
		\multirow{3}*{$d=7$, $N\geq 6$, $N\neq 29,30$}& \multirow{2}*{$\moa(28,7^12^t,2)$, $3\leq t\leq 12$} & $d=7$, $3\leq t \leq 5$, $N\geq 7$, $N\neq 30$, $31$ \\
		& \multirow{2}*{$\oa(28,2^t,2)$, $6\leq t\leq 27$} &$d=7$, $6\leq t \leq 12$, $N\geq 6$, $N\neq 30$\\
		& &$d=7$, $13\leq t \leq 27$, $N\geq 6$, $N\neq 29,30$\\	
		\hline	
	\end{tabular}
\end{table}

By using the same method as Example~\ref{ex:repl}, we have Table~\ref{Table:ct1ct2}.
One can refer to  Refs.~\cite{hedayat1999orthogonal,Orthogaonaltable} for more constructions of simple MOAs. 

\subsection{A construction of $3$-uniform states in heterogeneous systems}
In ~\cite{goyeneche2016multipartite}, the authors wondered whether there exist
some examples of $3$-uniform states in heterogeneous systems. Now, let us deal with this problem.

 Let $S$ be an abelian group of order $d$, and $S^k$ denote the abelian group of order $d^k$ consisting of all $k$-tuples of elements from $S$ with the usual vector addition as the binary operation. Let $S_0^k=\{(x_1,x_2,\ldots,x_k): x_1=x_2=\cdots=x_k\in S\}$, then $S_0^k$ is a subgroup of $S^k$ of order $d$,
and we will denote its cosets by $S_i^k$, $i=0,1,\ldots,d^{k-1}-1$. An $s\times N$ matrix $D$ based on $S$ is a \emph{difference scheme of strength $k$} \cite[Chapter 6]{hedayat1999orthogonal} if  for every $s \times k$ submatrix, by viewing each row as a coset representative,  each coset $S_i^k$, $i=0,1,\ldots,d^{k-1}-1$, is represented equally often. Denote it by $D_k(s,N,d)$. When $k=2$, one can check that $D_k(s,N,d)$ is $D(s,N,d)$.

 In the following we will show that a \emph{Hadamard matrix} can be taken as a difference scheme of strength 2 and 3. A Hadamard matrix $H_m$ is an $m\times m$ matrix  with entries $+1$'s and $-1$'s, which satisfies $H_m^{T}H_m=mI_m$ \cite{hedayat1999orthogonal}. If we replace $-1$ with $0$ in $H_m$, then $H_m$ is a GHM $D(m,m,2)$ over $\bbZ_2$, and it is also a $D_3(m,m,2)$ over $\bbZ_2$ \cite{chenguangzhou}. See Refs.~\cite{hedayat1999orthogonal,Differ} for more constructions of Hadamard matrices. The following is an  $H_4$,
 \begin{equation}
 H_4=\begin{pmatrix}
 1& 1 &1 &1\\
 1& 1 &0 &0\\
 1& 0 &1 &0\\
 1& 0 &0 &1 	
 \end{pmatrix}.
 \end{equation}

It is known that $D_3(s,N,d)$ can be used to construct MOAs with strength $3$.

\begin{construction}\label{ct:d3}
 Assume $(A_1,A_2)$ is an $\moa(r,d^{n_1}2^{n_2},3)$ with $\md(A_2)=b$, where $A_1$ is an $r\times n_1$ matrix, and $A_2$ is an $r\times n_2$ matrix.   Assume there also exists a $D_3(m,N,d)$ $D$ and  a Hadamard matrix $H_m$, then $(A_1\oplus D, A_2\oplus H_m)$ is an $\moa(rm,d^{n_1N}2^{n_2m},3)$ with $\md(A_2\oplus H_m)= \min\{\frac{m}{2}n_2,mb\}$.
\end{construction}
\begin{proof}
	The  construction of $\moa(rm,d^{n_1N}2^{n_2m},3)$ is from Ref.~\cite{chenguangzhou}. We only need to show $\md(A_2\oplus H_m)= \min\{\frac{m}{2}n_2,mb\}$. Since a Hadamard matrix is also a $\ghm$, we can obtain  $\md(A_2\oplus H_m)=\min \{\frac{m}{2}n_2,mb\}$ for the same discussion as Construction~\ref{ct:diffenencesch}.
\end{proof}
\vspace{0.5cm}

Now we give a family of $3$-uniform states in heterogeneous systems by Construction~\ref{ct:d3}, which answer the question posed in  \cite{goyeneche2016multipartite}. 
\begin{proposition}\label{pro:3uni}
	\begin{enumerate}[(i)]
		\item For any $n\geq 1$, there exists a 3-uniform state in $(\bbC^3)^{\otimes N}\otimes(\bbC^2)^{\otimes t}$ for any $N$ and $t$ with  $0\leq N\leq 4^n$ and $7\times 36^n +4\leq t\leq 9\times 36^n$.
		\item For any $n\geq 1$, there exists a 3-uniform state in $(\bbC^5)^{\otimes N}\otimes(\bbC^2)^{\otimes t}$ for any $N$ and $t$  with $0\leq N\leq 6^n$ and $5\times100^n+4\leq t\leq 6\times 100^n$.
	 	\end{enumerate}
\end{proposition}
	    \begin{proof}
	    (i) Since there exists a matrix $(A_1,A_2)$ which is an $\moa(48,3^12^9,3)$ with $\md(A_2)=2$, a $D_3(36,4,3)$ $D$, and a Hadamard matrix $H_{36}$  \cite{brouwer2006orthogonal,chenguangzhou},  $(A_1\oplus D,A_2\oplus H_{36})$ is an $\moa(48\times 36,3^42^{9\times 36},3)$ with $\md(A_2\oplus H_{36})=2\times 36$ by Construction~\ref{ct:d3}. Repeating this process $n$ times, $(A_1\oplus D\oplus\cdots \oplus D, A_2\oplus H_{36}\oplus \cdots\oplus H_{36})$ is an $\moa(48\times 36^n,3^{4^n}2^{9\times 36^n},3)$ with $\md(A_2\oplus H_{36}\oplus \cdots\oplus H_{36})=2\times 36^n$. There exists an $\irm(r,3^{N}2^{t},3)$ for any $0\leq N\leq 4^n$  and $7\times 36^n +4\leq t\leq 9\times 36^n$ by Lemma~\ref{Lemma:deletec} (The value of $N$ doesn't affect the irredundancy of the $\irm(r,3^{N}2^{t},3)$). Thus for any $n\geq 1$, there exists a 3-uniform state in $(\bbC^3)^{\otimes N}\otimes(\bbC^2)^{\otimes t}$ for any $0\leq N\leq 4^n$ and $7\times 36^n +4\leq t\leq 9\times 36^n$ by Proposition~\ref{pp:moaandkuniform}.
	
	 	(ii) Since there exists a matrix $(A_1,A_2)$ which is $\moa(40,5^12^6,3)$ with $\md(A_2)=1$, a $D_3(100,6,5)$, and a Hadamard matrix $H_{100}$ \cite{brouwer2006orthogonal,chenguangzhou},   we can obtain a 3-uniform state in $(\bbC^5)^{\otimes N}\otimes(\bbC^2)^{\otimes t}$ for any $0\leq N\leq 6^n$, $5\times100^n+4\leq t\leq 6\times 100^n$ and $n\geq 1$  for  the same discussion as (i).
 \end{proof}
\vspace{0.5cm}

 For more constructions of $\moa$s with strength $3$ and difference schemes of strength $3$,  see Refs.  \cite{brouwer2006orthogonal,nguyen2008some,wang2019construction,chenguangzhou}.
 
\subsection{Constructions of $(k-1)$-uniform states from $k$-uniform states.}\label{sec:ktok-1}

In this subsection, we give two methods of constructing $(k-1)$-uniform states in heterogeneous systems from $k$-uniform states in heterogeneous systems. The first method is inspired by Refs.~\cite{goyeneche2014genuinely,li2019k}.

\begin{proposition}\label{pro:projective}
	Assume $\ket{\psi}$ is a $k$-uniform state in $\bbC^{d_1}\otimes\bbC^{d_{2}}\otimes\cdots\otimes\bbC^{d_{N}}$, and it is written in the computational basis. If Alice performs the projective measurement with projectors $\{\ketbra{i}{i}\}_{i=0}^{d_1-1}$ on the first system,  then we obtain a $(k-1)$-uniform state in $\bbC^{d_{2}}\otimes\cdots\otimes\bbC^{d_{N}}$ for each projector. Moreover, these $(k-1)$-uniform states are pairwise orthogonal.
\end{proposition}

See Appendix~\ref{Appendix:E} for the proof of Proposition~\ref{pro:projective}. Recall that Eq.~(\ref{eq:42222}) gives a $2$-uniform state $\ket{\psi}$ in $\bbC^4\otimes(\bbC^2)^{\otimes 4}$. If Alice performs the projective measurement with projectors $\{\ketbra{i}{i}\}_{i=0}^{3}$ on the first system,  then we can obtain four pairwise orthogonal $1$-uniform states in $(\bbC^2)^{\otimes 4}$ by Proposition~\ref{pro:projective}: $\ket{\psi_0}=\frac{1}{\sqrt{2}}(\ket{0000}+\ket{1111})$, $\ket{\psi_1}=\frac{1}{\sqrt{2}}(\ket{0011}+\ket{1100})$, $\ket{\psi_2}=\frac{1}{\sqrt{2}}(\ket{0101}+\ket{1010})$, $\ket{\psi_3}=\frac{1}{\sqrt{2}}(\ket{0110}+\ket{1001})$. If Alice performs the projective measurement with projectors $\{\ketbra{i}{i}\}_{i=0}^{1}$ on the fifth system,  then we obtain two  orthogonal $1$-uniform states in $\bbC^4\otimes(\bbC^2)^{\otimes 3}$ by Proposition~\ref{pro:projective}:
$\ket{\psi'_0}=\frac{1}{2}(\ket{0000}+\ket{1110}+\ket{2101}+\ket{3011})$, $\ket{\psi'_1}=\frac{1}{2}(\ket{0111}+\ket{1001}+\ket{2010}+\ket{3100})$.

Now we introduce the second method.

\begin{proposition}
	\label{Pro:k+1-k}
	If there exists a $k$-uniform state in $\bbC^{d_1}\otimes\bbC^{d_2}\otimes\bbC^{d_3}\otimes\cdots\otimes\bbC^{d_N}$,  then there exists a $(k-1)$-uniform state in
	$\bbC^{d_1d_2}\otimes\bbC^{d_3}\otimes\cdots\otimes\bbC^{d_N}$.
\end{proposition}
\begin{proof}
	Assume $\ket{\psi}$ is a  $k$-uniform state in $\bbC^{d_1}\otimes\bbC^{d_2}\otimes\bbC^{d_3}\otimes\cdots\otimes\bbC^{d_n}$. If we consider the vectors in $\bbC^{d_1}\otimes\bbC^{d_2}$ as the vectors in $\bbC^{d_1d_2}$, then $\ket{\psi}$ is a $(k-1)$-uniform state in
	$\bbC^{d_1d_2}\otimes\bbC^{d_3}\otimes\cdots\otimes\bbC^{d_N}$. We can check that  if $\{i_1,i_2,\ldots,i_{k-1}\}\in\{2,3,\ldots,N-1\}$,
	\begin{equation}
	\tr_{\{i_1,i_2,\ldots,i_{k-1}\}^c}\ketbra{\psi}{\psi}=\frac{1}{d_{i_1+1}d_{i_2+1}\cdots d_{i_{k-1}+1}}I_{d_{i_1+1}d_{i_2+1}\cdots d_{i_{k-1}+1}}.
	\end{equation}	
	If $\{i_2,\ldots,i_{k-1}\}\in\{2,3,\ldots,N-1\}$,
	\begin{equation}
	\tr_{\{1,i_2,\ldots,i_{k-1}\}^c}\ketbra{\psi}{\psi}=\frac{1}{d_1d_2d_{i_2+1}\cdots d_{i_{k-1}+1}}I_{d_1d_2d_{i_2+1}\cdots d_{i_{k-1}+1}}.
	\end{equation}	
\end{proof}
\vspace{0.5cm}	

Specially, we can construct $(k-1)$-uniform states in heterogeneous systems from $k$-uniform states in homogeneous systems by Proposition~\ref{Pro:k+1-k}.
For example, since there exists an AME states in $(\bbC^{3})^{\otimes 10}$ \cite{AMEtable},  there exists a $4$-uniform state in $\bbC^{9}\otimes (\bbC^{3})^{\otimes 8}$, and it is also an AME states in $\bbC^{9}\otimes (\bbC^{3})^{\otimes 8}$. 
The methods in Proposition~\ref{pro:projective} and \ref{Pro:k+1-k}  are also useful  for quantum information masking in the multipartite
scenario \cite{modi2018masking}.

\section{AME states in heterogeneous systems}\label{sec:AMEkuniform}
\label{sec:AME}
 In this section, we show some results on the existence and nonexistence of AME states  in heterogeneous systems. The existence of AME states in homogeneous systems have been attracting much  attention \cite{rains1999quantum,scott2004multipartite,huber2017absolutely,huber2018bounds,horodecki2020five}. But there are few existence results of AME states in heterogeneous systems. AME states in  tripartite heterogeneous systems is investigated in \cite{shen2020absolutely}.
For an AME state in $\bbC^{d_1}\otimes\bbC^{d_2}\otimes\cdots\otimes\bbC^{d_N}$, the definition is the same as homogenous systems. It requires all reductions to $\fl{N}{2}$ parties are maximally mixed. This is the definition that we focused on in this paper. There is  another definition of AME states in $\bbC^{d_1}\otimes\bbC^{d_2}\otimes\cdots\otimes\bbC^{d_N}$, which requires every subsystem whose dimension is not
larger than that of its complement must be maximally mixed \cite{huber2018bounds}.

For an AME state in $\bbC^{d_1}\otimes\bbC^{d_2}\otimes\cdots\otimes\bbC^{d_N}$, if $d_1,d_2,\ldots,d_N$ are not all equal, then $N$ must be odd since $d_1\cdots d_{\fl{N}{2}}\leq d_{\fl{N}{2}+1}\cdots d_N$. There are some examples of AME states.  Eq.~(\ref{eq:42222}) gives an AME state in $\bbC^4\otimes(\bbC^2)^{\otimes 4}$. Since there exists an $\oa(16,4^5,2)$ with $\md=4$ \cite{hedayat1999orthogonal,pang2019two}, we can obtain an $\moa(16,4^42^2,2)$ with $\md\geq 4$ by splitting method. Further we have an $\irm(16,4^32^2,2)$ and an $\irm(16,4^42^1,2)$ by Lemma~\ref{Lemma:deletec}. It means that there are  AME states in $(\bbC^4)^{\otimes 3}\otimes(\bbC^2)^{\otimes 2}$ and $(\bbC^4)^{\otimes 4}\otimes\bbC^2$.  We have the following corollary from Proposition~\ref{Pro:k+1-k}.

\begin{corollary}\label{cor:AME}
	If there exists an AME states in $(\bbC^{d})^{\otimes N}$, $N$ is even, then there exists an AME state in $\bbC^{d^2}\otimes(\bbC^{d})^{\otimes N-2}$.
\end{corollary}

Now, we consider the nonexistence of AME states in heterogeneous systems.
In \cite{huber2018bounds}, they showed the nonexistence of some AME states in homogeneous systems by shadow inequalities. Inspired by this idea,  we can also show the nonexistence of some AME states in  heterogeneous systems by shadow inequalities.
Assume an AME state exists in $\bbC^{d_1}\otimes \bbC^{d_2}\otimes\cdots\otimes\bbC^{d_N}$, where $N$ is odd. The main idea is to show the following necessary conditions do not hold:
\begin{equation}\label{eq:shadow}
S_{j}=\sum_{k=0}^NK_{N-j}(k;N)A_k'\geq 0
\end{equation}
for all $0\leq j\leq N$, where $K_{N-j}(k;N)=\sum_{\alpha}(-1)^{\alpha}\binom{N-k}{N-j-\alpha}\binom{k}{\alpha}$,
$A_0'=1$, $A_k'=\sum_{\{i_1,i_2,\ldots,i_k\}\subset\{1,2,\ldots,N\}}\frac{1}{d_{i_1}d_{i_2}\cdots d_{i_k}}$ for $1\leq k\leq \frac{N-1}{2}$, and $A_k'=A_{N-k}'$. Specially, if $d_1=d_2=\cdots=d_N=d$, then $A_k'=\binom{N}{k}d^{-\min(k,N-k)}$.

Now, we can give the nonexistence of some AME states in $\bbC^{d_1}\otimes \bbC^{d_2}\otimes\ldots\otimes\bbC^{d_N}$ by Eq.~(\ref{eq:shadow}). See the following lemma and Table~\ref{Table:nonAME}. Table~\ref{Table:nonAME} is obtained by computer.

\begin{table}
	\renewcommand\arraystretch{1.7}	
	\caption{The nonexistence of AME states in heterogeneous systems.}
	\label{Table:nonAME}
	\centering
	\renewcommand\tabcolsep{6.0pt}
	\begin{tabular}{ccc}
		\hline
		9-parties& 11-parties &13-parties \\
		\hline
		$\bbC^3\otimes(\bbC^2)^{\otimes 8}$& $\bbC^3\otimes(\bbC^2)^{\otimes 10}$	&$(\bbC^3)^{\otimes n}\otimes(\bbC^2)^{\otimes (13-n)}$, $n=1,11,12$\\
		$(\bbC^3)^{\otimes 7}\otimes(\bbC^2)^{\otimes 2}$& $\bbC^4\otimes(\bbC^2)^{\otimes 10}$ & $\bbC^4\otimes(\bbC^2)^{\otimes 12}$\\
		$\bbC^4\otimes(\bbC^2)^{\otimes 8}$& &$(\bbC^4)^{\otimes n}\otimes(\bbC^3)^{\otimes (13-n)}$, $n=1,2,3$\\
		$\bbC^4\otimes(\bbC^3)^{\otimes 6}\otimes(\bbC^2)^{\otimes 2}$& &$(\bbC^5)^{\otimes n}\otimes(\bbC^3)^{\otimes (13-n)}$, $n=1,2$\\
		& &$\bbC^d\otimes(\bbC^3)^{\otimes 12}$, $d=6,7,8,9$\\
		& &$(\bbC^4)^{\otimes n_1}\otimes(\bbC^3)^{\otimes n_2}\otimes(\bbC^2)^{\otimes (13-n_1-n_2)}$,  $(n_1,n_2)=(1,10),(1,11),(2,10)$\\
		& &$(\bbC^5)^{\otimes n_1}\otimes(\bbC^4)^{\otimes n_2}\otimes(\bbC^3)^{\otimes (13-n_1-n_2)}$,  $(n_1,n_2)=(1,1),(1,2)$\\
		& &$\bbC^5\otimes(\bbC^3)^{\otimes 11}\otimes\bbC^2$\\
		& &$\bbC^6\otimes\bbC^4\otimes(\bbC^3)^{\otimes 11}$\\
		\hline
	\end{tabular}
\end{table}

\begin{lemma}\label{lem:ame3128}
AME states in $\bbC^3\otimes(\bbC^2)^{\otimes 8}$ do not exist.
\end{lemma}
\begin{proof}
Since $d_1=3$ and $d_i=2$ for $2\leq i\leq 8$,
$A_0'=1$, $A_1'=\frac{13}{3}$, $A_2'=\frac{25}{3}$, $A_3'=\frac{28}{3}$, $A_4'=\frac{161}{24}$, $A_j'=A_{9-j}'$ for $5\leq j\leq 9$. We have $K_{8}(0,9)=9$, $K_{8}(1,9)=-7$, $K_{8}(2,9)=5$, $K_{8}(3,9)=-3$, $K_{8}(4,9)=1$, $K_{8}(k,9)=K_{8}(9-k,9)$  for $5\leq k\leq 9$. We can calculate that
\begin{equation}
S_1=\sum_{k=0}^{9}K_{8}(k,9)A_k'=-\frac{23}{12}<0.
\end{equation}
Thus AME states in $\bbC^3\otimes(\bbC^2)^{\otimes 8}$ do not exist.
\end{proof}

In next section, we consider the applicaitions $k$-uniform states.

\section{Applications}\label{sec:twoapp}

In this section, we give some applications of $k$-uniform states in heterogeneous systems.  We construct an orthogonal basis consisting of $k$-uniform states in Proposition~\ref{Pro:k-basis}, and show that some $k$-uniform bases are locally irreducible in Proposition~\ref{pro:local} and Corollary~\ref{cor:ghzbasis}.

\subsection{$k$-uniform bases in heterogeneous systems}
In this subsection, we construct an orthogonal basis consisting of $k$-uniform states in a heterogeneous system.
In a homogeneous system $(\bbC^d)^{\otimes N}$,  the \emph{support} of a state is the number of nonzero coefficients, when it is written in the computational
basis.  Since all the reductions to
$k$ parties of a $k$-uniform state are maximally mixed, the support of a $k$-uniform state is at least $d^k$. A $k$-uniform state is called a  $k$-uniform state with \emph{minimum support} in $(\bbC^d)^{\otimes N}$ if the support is $d^k$ \cite{goyeneche2015absolutely}. We can generalize this definition for heterogeneous systems. In a heterogeneous system $\bbC^{d_1}\otimes\bbC^{d_2}\otimes\cdots\otimes\bbC^{d_N}$, $d_1\geq d_2\geq \cdots\geq d_N$, the support of a $k$-uniform state is at least $d_1d_2\cdots d_k$, due to that all the reductions to
first $k$ parties of the $k$-uniform state are maximally mixed.  A $k$-uniform state is called a  $k$-uniform state with minimum support  in $\bbC^{d_1}\otimes\bbC^{d_2}\otimes\cdots\otimes\bbC^{d_N}$ if the support is  $d_1d_2\cdots d_k$. For example, Eq.~(\ref{eq:42222}) gives a 2-uniform state with minimum support   in $\bbC^4\otimes(\bbC^2)^{\otimes 4}$.

In \cite{li2019k,raissi2019constructing}, they showed that if there exists a  $k$-uniform state with minimum support in $(\bbC^d)^{\otimes N}$, then there exists an orthogonal basis consisting of $k$-uniform states with minimum support  in  $(\bbC^d)^{\otimes N}$. This is also true for heterogeneous systems.

Assume $\ket{\psi}$ is a $k$-uniform state with minimum support in $\bbC^{d_1}\otimes\bbC^{d_2}\otimes\cdots\otimes\bbC^{d_N}$.
Define the generalized  Pauli operators,
\begin{align*}
&Z_i\ket{j}=w_i^j\ket{j},  &w_i=e^{\frac{2\pi\sqrt{-1}}{d_i}};\\
&X_i\ket{j}=\ket{j\oplus_i 1}, &j\oplus_i 1=j+1\pmod {d_i}.
\end{align*}
Define
\begin{equation}
U(\bv)=Z_1^{v_1}\otimes\cdots\otimes Z_k^{v_k}\otimes X_{k+1}^{v_{k+1}}\otimes\cdots\otimes X_{N}^{v_{N}},
\end{equation}
where $\bv=(v_1,v_2,\ldots,v_N)\in \bbZ_{d_1}\times\bbZ_{d_2}\times\cdots\times\bbZ_{d_N}$. We have the following proposition.

\begin{proposition}\label{Pro:k-basis}
	If there exists a  $k$-uniform state $\ket{\psi}$ with minimum support in $\bbC^{d_1}\otimes\bbC^{d_2}\otimes\cdots\otimes\bbC^{d_N}$,  then $\mathbf{\cB}=\{U(\bv)\ket{\psi}: \bv\in \bbZ_{d_1}\times\bbZ_{d_2}\times\cdots\times\bbZ_{d_N}\}$ is an orthogonal basis consisting of $k$-uniform states with minimum support in $\bbC^{d_1}\otimes\bbC^{d_2}\otimes\cdots\otimes\bbC^{d_N}$.
\end{proposition}

See Appendix~\ref{Appendix:B} for the proof of Proposition~\ref{Pro:k-basis}.

Remark that if there exists a  $k$-uniform state with minimum support in $\bbC^{d_1}\otimes\bbC^{d_2}\otimes\cdots\otimes\bbC^{d_N}$, then  $d_{i_1}d_{i_2}\cdots d_{i_k}$ must divide $d_1d_2\cdots d_k$, for any $\{i_1,i_2,\ldots,i_k\}\subset \{1,2,\ldots,N-1\}$.

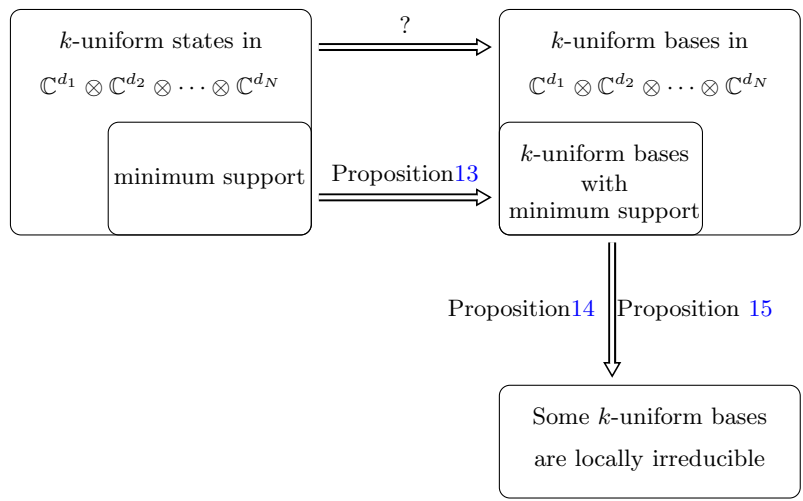
\begin{figure}[h]
	\begin{tikzpicture}
	\draw[rounded corners] (-5,0) rectangle (-1,3);
	\draw (-3,2.6) node []{$k$-uniform states in};
	\draw (-3,2) node []{$\bbC^{d_1}\otimes\bbC^{d_2}\otimes\cdots\otimes\bbC^{d_N}$};
	\draw[rounded corners]  (-3.7,0) rectangle (-1,1.5);
	\draw (-2.35,0.75) node []{minimum support};

	\draw[rounded corners] (1.5,0) rectangle (5.5,3);
	\draw (3.5,2.6) node []{$k$-uniform bases in};
	\draw (3.5,2) node []{$\bbC^{d_1}\otimes\bbC^{d_2}\otimes\cdots\otimes\bbC^{d_N}$};
	\draw[rounded corners] (1.5,0) rectangle (4.2,1.5);
	\draw (2.9,1.1) node []{$k$-uniform bases};
	\draw (2.9,0.7) node []{with};
	\draw (2.9,0.3) node []{minimum support};

	\draw[rounded corners] (1.5,-3.5) rectangle (5.5,-2);
	\draw (3.5,-2.4) node []{Some $k$-uniform bases };
	\draw (3.5,-3) node []{ are locally irreducible};

	\draw [vecArrow](-0.9,2.5)--(1.4,2.5);
	\draw [vecArrow](-0.9,0.5)--(1.4,0.5);
	\draw [vecArrow](3,-0.1)--(3,-1.9);

	\draw (0.25,2.8) node []{?};
	\draw (0.25,0.8) node []{Proposition\ref{Pro:k-basis}};
	\draw (1.8,-1) node []{Proposition\ref{pro:local}};
	\draw (4.1,-1) node []{Proposition~\ref{cor:ghzbasis}};
	
	\end{tikzpicture}
	\caption{The applications that we obtained in Sec.~\ref{sec:twoapp}. A $k$-uniform state  is called a   $k$-uniform state with minimum support  in $\bbC^{d_1}\otimes\bbC^{d_2}\otimes\cdots\otimes\bbC^{d_N}$, if it has $d_1d_2\cdots d_k$ nonzero coefficients when it is written in the computational basis. Note that ``?'' means that we don't know whether we can construct a $k$-uniform bases  from a  $k$-uniform state with non-minimum support in  $\bbC^{d_1}\otimes\bbC^{d_2}\otimes\cdots\otimes\bbC^{d_N}$. }\label{Fig:mini}
\end{figure}

\subsection{Local indistinguishability of $k$-uniform bases in heterogeneous systems}
Now, we discuss the local indistinguishability of $k$-uniform bases, this shows quantum nonlocality with  entanglement. A measurement performed to distinguish a set of mutually orthogonal states is called an orthogonality-preserving measurement if after the measurement the states
remain mutually orthogonal.  A measurement is nontrivial if not all the
POVM elements are proportional to the identity operator.
Otherwise, the measurement is trivial. A set of orthogonal  states is \emph{locally irreducible} if it is not possible to eliminate one or more  states from
the set by nontrivial orthogonality-preserving local measurements \cite{halder2019strong}. Local irreducibility sufficiently ensures
local indistinguishability, while the converse is not ture. Recall that  Eq.~(\ref{eq:42222}) gives a $2$-uniform state  with minimum support in $\bbC^4\otimes(\bbC^2)^{\otimes 4}$. Then we can obtain an orthogonal basis 	$\cB=\{U(\bv)\ket{\psi}:\bv\in\bbZ_4\times\bbZ_2\times\bbZ_2\times\bbZ_2\times\bbZ_2 \}$ in $\bbC^4\otimes(\bbC^2)^{\otimes 4}$ by Proposition~\ref{Pro:k-basis}.  We have the following proposition.
\begin{proposition}\label{pro:local}
	The  basis $\cB$ consisting of $2$-uniform states with minimum support in $\bbC^4\otimes(\bbC^2)^{\otimes 4}$ is locally irreducible.
\end{proposition}

See Appendix~\ref{Appendix:C} for the proof of Proposition~\ref{pro:local}. Since the states of $\cB$ are all AMEs, the local indistinguishability of
$\cB$ shows quantum nonlocality with maximum entanglement. In \cite{halder2019strong}, the authors showed that the $N$-qubit $\ghz$ bases are locally irreducible.  More generally, we can consider the  $N$-qudit $\ghz$ bases. Since
\begin{equation}
\ket{\ghz}=\frac{1}{\sqrt{d}}\sum_{j=0}^{d-1}\ket{j}^{\otimes N}
\end{equation}
is a $1$-uniform state with minimum support in $(\bbC^{d})^{\otimes N}$, we can also obtain an orthogonal basis $\mathbf{\cB_1}=\{U(\bv)\ket{\ghz}: \bv\in \bbZ_{d}\times\bbZ_{d}\times\cdots\times\bbZ_{d}\}$ in $(\bbC^{d})^{\otimes N}$. For the same discussion as Proposition~\ref{pro:local}, we have the following proposition.
\begin{proposition}\label{cor:ghzbasis}
		The  basis $\cB_1$ consisting of $1$-uniform states with minimum support in $(\bbC^{d})^{\otimes N}$ is locally irreducible.
\end{proposition}

Since those bases are locally indistinguishable, one can also consider to distinguish  those bases assisted by entanglement as
a nonlocal resource. Further we can consider the local irreducibility of other $k$-uniform basis.  See Figure~\ref{Fig:mini} for all the applications in Section~\ref{sec:twoapp}.

\section{Conclusion}
\label{sec:conclusionpart}
In this paper, we have investigated $k$-uniform states in heterogeneous systems.  We have given the connections between  $\moa$s with certain Hamming distance, $\irm$s, and $k$-uniform states.  We have given two constructions of  $2$-uniform states in heterogeneous systems. We have also given a construction of $3$-uniform states in heterogeneous systems, which solves a question in \cite{goyeneche2016multipartite}. Moreover, we have shown two methods of generating $(k-1)$-uniform states from $k$-uniform states. We have presented some new results on existence and nonexistence of AME states in heterogeneous systems. For applications, we have presented an orthogonal basis consisting of  $k$-uniform states with minimum support, and  shown that some $k$-uniform bases are locally irreducible.  An open problem is to find the unknown cases of $k$-uniform states in Table~\ref{Table:k-unfiormresults}.  Another problem is whether any $k$-uniform state is locally unitarily equivalent to a  $k$-uniform states with real coefficients.

\section*{Acknowledgments}
\label{sec:ack}	
FS and XZ were supported by NSFC under Grant No. 11771419,  the Fundamental Research Funds for the Central Universities, and Anhui Initiative in Quantum Information Technologies under Grant No. AHY150200. LC and YS were supported by the  NNSF of China (Grant No. 11871089), and the Fundamental Research Funds for the Central Universities (Grant No. ZG216S2005).

\appendix

\section{The proof of Construction~\ref{ct:diffenencesch}}
\label{Appendix:A}
\begin{proof}
	The construction of $\moa(r\lambda d,(\lambda d)^1d^{N\lambda d},2)$ is from \cite{hedayat1999orthogonal}. We only need to show $\md(A_1')= \min\{\lambda(d-1)N+1, \lambda db\}$.  Assume $A_1=(a_{i,j})_{1\leq i\leq r,1\leq j\leq N}$, $A_2=(f_{i,j})_{1\leq i\leq \lambda d,1\leq j\leq \lambda d}$,
	\begin{equation}
	A_1'=(\bm^T, A_1\oplus A_2)=
	\begin{pmatrix}
	\bm_1 &a_{1,1}+A_2 & a_{1,2}+A_2 & \cdots & a_{1,N}+A_2 \\
	\bm_2 &a_{2,1}+A_2 & a_{2,2}+A_2 & \cdots & a_{2,N}+A_2 \\
	\vdots &\vdots & \vdots &\ddots& \ldots  \\
	\bm_r&a_{r,1}+A_2 & a_{r,2}+A_2 & \cdots & a_{r,N}+A_2 	
	\end{pmatrix}=\begin{pmatrix}
	\bm_1 &B_{1,1}  \\
	\bm_2  &B_{2,1} \\
	\vdots &\vdots  \\
	\bm_r  & B_{r,1} 	
	\end{pmatrix},
	\end{equation}
	where $B_{i,1}=(
	a_{i,1}+A_2,a_{i,2}+A_2,\ldots,a_{i,N}+A_2)$, $\bm_i=(0,1,\ldots,\lambda d-1)^T$, $1\leq i\leq r$. Let  $\bv_1'=(t,\bv_1)$ and$\bv_2'=(t',\bv_2)$ be two row vectors of $A_1'$, where $\bv_1\in B_{i,1},\bv_2\in B_{j,1}$. There are two cases.
	
	If $i=j$, then we can assume
	$$
	\bv_1=(
	a_{i,1}+f_{k,1}, a_{i,1}+f_{k,2},\ldots,a_{i,1}+f_{k,\lambda d},\ldots ,a_{i,N}+f_{k,1} ,\ldots ,a_{i,N}+f_{k,\lambda d}),
	$$$$
	\bv_2=(a_{i,1}+f_{\ell,1} ,a_{i,1}+f_{\ell,2} ,\ldots ,a_{i,1}+f_{\ell,\lambda d},\ldots ,a_{i,N}+f_{\ell,1}  ,\ldots ,a_{i,N}+f_{\ell,\lambda d}),
	$$
	where $k\neq \ell$. We have $t=k-1\neq\ell-1=t'$,  $\hd(\bv_1,\bv_2)=\hd(\bv_1-\bv_1,\bv_2-\bv_1)$.
	Since $A_2$ is a $\ghm$,  $\{0,1\ldots,d-1\}$ appears $\lambda N$ times in $\bv_2-\bv_1$. Thus $\hd(\bv_1,\bv_2)=\lambda(d-1)N$. Further $\hd(\bv_1',\bv_2')=\lambda(d-1)N+1$  due to $t\neq t'$.
	
	If $i\neq j$, then we can assume
	$$
	\bv_1=(a_{i,1}+f_{k,1} ,a_{i,1}+f_{k,2} ,\ldots ,a_{i,1}+f_{k,\lambda d},\ldots ,a_{i,N}+f_{k,1} ,\ldots ,a_{i,N}+f_{k,\lambda d}),
	$$$$
	\bv_2=(a_{j,1}+f_{\ell,1} ,a_{j,1}+f_{\ell,2} ,\ldots ,a_{j,1}+f_{\ell,\lambda d},\ldots ,a_{j,N}+f_{\ell,1}  ,\ldots ,a_{j,N}+f_{\ell,\lambda d}).
	$$
	If $k=\ell$, then $t=k-1=\ell-1=t'$. We have $\hd(\bv_1',\bv_2')= \hd(\bv_1,\bv_2)=\lambda db$. If $k\neq \ell$, then $t=k-1\neq\ell-1=t'$. We denote
	$$
	\be=(f_{k,1} ,f_{k,2} ,\ldots ,f_{k,\lambda d},\ldots ,f_{k,1}  ,\ldots ,f_{k,\lambda d}
	).
	$$
	We have $\hd(\bv_1,\bv_2)=\hd(\bv_1-\be,\bv_2-\be)$, where
	$$
	\bv_1-\be=(
	a_{i,1} ,a_{i,1} ,\ldots ,a_{i,1},\ldots ,a_{i,N} ,\ldots ,a_{i,N}),
	$$
	\begin{align*}
	\bv_2-\be=(&a_{j,1}+f_{\ell,1}-f_{k,1},  a_{j,1}+f_{\ell,2}-f_{k,2}, \ldots, a_{j,1}+f_{\ell,\lambda d}-f_{k,\lambda d}, \ldots \\
	&a_{j,N}+f_{\ell,1}-f_{k,1}, a_{j,N}+f_{\ell,2}-f_{k,2}, \ldots , a_{j,N}+f_{\ell,\lambda d}-f_{k,\lambda d}).\\
	\end{align*}
	Through a permutation of $\bv_2-\be$, we have
	\begin{align*}
	(\bv_2-\be)'=(&a_{j,1}+0 , a_{j,1}+1 , \ldots , a_{j,1}+d-1, \ldots, a_{j,1}+0 , a_{j,1}+1 , \ldots , &\\
	&a_{j,1}+d-1, \ldots , a_{j,N}+0 , a_{j,N}+1 , \ldots , a_{j,N}+d-1 , \ldots , a_{j,N}+0 , \\
	&a_{j,N}+1 , \ldots ,a_{j,N}+d-1),\\
	\end{align*}
	and $\hd(\bv_1-\be,(\bv_2-\be)')=\hd(\bv_1-\be,\bv_2-\be)$. Since $\hd((a_{i,1} , a_{i,1}, \ldots,a_{i,1}),(a_{j,1}+0 , a_{j,1}+1 , \ldots , a_{j,1}+d-1))=d-1$,  $\hd(\bv_1,\bv_2)=\hd(\bv_1-\be,(\bv_2-\be)')=\lambda(d-1)N$ and $\hd(\bv_1',\bv_2')=\lambda(d-1)N+1$.
	
	In conclusion, we have  $\md(A_1')= \min\{\lambda(d-1)N+1, \lambda db\}$.
\end{proof}

\section{The proof of Theorem~\ref{Thm:dddd2}}
\label{Appendix:D}
\begin{proof}
	(i)
	If $d$ is a prime power, then there exists a $\ghm$ $D(4d^n,4d^n,d)$ for any $n\geq 1$ \cite[Chapter 6]{hedayat1999orthogonal}.  Let $A_1=(
	0,1,\ldots,d-1)^T$ and $A_2$ be a $D(4d^n,4d^n,d)$ in Construction~\ref{ct:diffenencesch}, we obtain an $\moa(4d^{n+1},(4d^n)^1d^{4d^n},2)$  with $\md=4d^{n-1}(d-1)+1$. We further obtain an $\moa(4d^{n+1},d^{4d^n+n}2^2,2)$  with $\md\geq 4d^{n-1}(d-1)+1$  by splitting method.  Then there exists an $\irm(4d^{n+1},d^{N}2^2,2)$ for any
	$4d^{n-1}+n+2\leq N\leq 4d^n+n$ and $n\geq 1$ by Lemma~\ref{Lemma:deletec}. We only need to consider the cases when $N=4d^n+n+1,4d^n+n+2$ for $n\geq 1$.
	
	There exists an $\oa(d^n,d^{\frac{d^n-1}{d-1}},2)$ with $\md=d^{n-1}$ for any $n\geq 2$  \cite{pang2019two}. Let $A_1$ be an $\oa(d^n,\frac{d^n-1}{d-1},d,2)$ and $A_2$ be a $D(4d,4d,d)$  in Construction~\ref{ct:diffenencesch}, then we obtain an $\moa(4d^{n+1},(4d)^1d^{\frac{4d(d^n-1)}{d-1}},2)$ with $\md=4d^n-3$, and further an $\moa(4d^{n+1},d^{\frac{4d(d^n-1)}{d-1}+1}2^2,2)$ with $\md\geq 4d^n-3$  by splitting method. Then there exists an $\irm(4d^{n+1},d^{N'}2^2,2)$ for any $\frac{4d(d^n-1)}{d-1}-4d^n+7\leq N'\leq \frac{4d(d^n-1)}{d-1}+1$ and $n\geq 2$ by Lemma~\ref{Lemma:deletec}. Since $\frac{4d(d^n-1)}{d-1}-4d^n+7\leq 4d^n+n+1$, $4d^n+n+2\leq \frac{4d(d^n-1)}{d-1}+1$ for $n\geq 2$,  there exists an $\irm(4d^{n+1},d^{4d^n+n+1}2^2,2)$ and an $\irm(4d^{n+1},d^{4d^n+n+2}2^2,2)$ for $n\geq 2$.
	
	Thus,  when $d$ is a prime power, there exists an $\irm(r,d^N2^2,2)$ for any $N\geq 7$ and $N\neq 4d+2,4d+3$, and there exists a $2$-uniform state in $(\bbC^d)^{\otimes N}\otimes\bbC^2\otimes\bbC^2$ for any  $N\geq 7$ and $N\neq 4d+2,4d+3$ by Proposition~\ref{pp:moaandkuniform}. When $d'$ is a prime power, there exists a $2$-uniform state in $(\bbC^{d'})^{\otimes N}\otimes\bbC^1\otimes\bbC^1$ for any $N\geq 5$ \cite{li2019k}. By the tensor of a  $2$-uniform state  in $(\bbC^d)^{\otimes N}\otimes\bbC^2\otimes\bbC^2$ and a $2$-uniform state in $(\bbC^{d'})^{\otimes N}\otimes\bbC^1\otimes\bbC^1$ \footnote{If there exists an $n$-partite $k$-uniform state $\ket{\psi}_{A_1A_2\ldots A_n}$ in $\bbC^{d_1}\otimes\bbC^{d_2}\otimes\cdots\bbC^{d_n}$, and an $n$-partite $k$-uniform state  $\ket{\phi}_{B_1B_2\ldots B_n}$ in $\bbC^{s_1}\otimes\bbC^{s_2}\otimes\cdots\bbC^{s_n}$, then $\ket{\varphi}_{(A_1B_1)(A_2B_2)\ldots(A_nB_n)}=\ket{\psi}_{A_1A_2\ldots A_n}\otimes\ket{\phi}_{B_1B_2\ldots B_n}$ is an $n$-partite $k$-uniform state in $\bbC^{d_1s_1}\otimes\bbC^{d_2s_2}\otimes\cdots\bbC^{d_ns_n}$ \cite{shen2020absolutely}.}, we obtain a 2-uniform state in $(\bbC^d)^{\otimes N}\otimes\bbC^2\otimes\bbC^2$ for any $d\geq 2$, $N\geq 7$ and $N\neq 4d+2,4d+3$.
	
	(ii)
	If $d$ is a prime power, there exists also a $\ghm$ $D(2d^n,2d^n,d)$ for any $n\geq 1$ \cite[Chapter 6]{hedayat1999orthogonal}.  When $d$ is a prime power, there exists an $\irm(r,d^N2^1,2)$ for any $N\geq 5$, $N\neq 2d+2,2d+3$ by the same discussion as (i). We only need to consider the cases when $N=2d+2,2d+3$. From the proof of (i), we know that there exists an $\irm(4d^2,d^{4d+1}2^2,2)$ with $\md\geq 4d-3$. Then there exists an  $\irm(4d^2,d^N2^1,2)$ for any $8\leq N\leq 4d+1$ by Lemma~\ref{Lemma:deletec}. Since $8\leq 2d+2, 2d+3 \leq 4d+1$ for $d\geq 3$, and there exists an $\iro(r,2^N2^1,2)$ for any $N\geq 5$ \cite{goyeneche2014genuinely},  then when $d$ is a prime power, we obtain an $\irm(r,d^N2^1,2)$ for any $N \geq 5$. Thus, there exists a $2$-uniform states in $(\bbC^d)^{\otimes N}\otimes\bbC^2$ for any $d\geq 2$ and $N\geq 5$ by the same discussion as (i).	
\end{proof}

\section{The proof of Proposition~\ref{pro:projective}}
\label{Appendix:E}
\begin{proof}
	Since $\ket{\psi}$ is a $k$-uniform state, we choose any $(k-1)$ subsystems which are different from the first system, namely, $i_2,i_3,\ldots,i_k$. Then
	\begin{equation}
	\ket{\psi}=\sum_{(j_1,j_{i_2},\ldots,j_{i_k})\in\bbZ_{d_1}\times\bbZ_{d_{i_2}}\times\ldots\times\bbZ_{d_{i_k}}}c_{j_1,j_{i_2},\ldots,j_{i_k}}\ket{j_1,j_{i_2},\ldots,j_{i_k}}\ket{\psi(j_1,j_{i_2},\ldots,j_{i_k})},
	\end{equation}
	where $|c_{j_1,j_{i_2},\ldots,j_{i_k}}|=\frac{1}{\sqrt{d_1d_{i_2}\cdots d_{i_k}}}$, and $\braket{\psi(j_1,j_{i_2},\ldots,j_{i_k})}{\psi(j_1',j_{i_2}',\ldots,j_{i_k}')}=\delta_{j_1,j_1'}\delta_{j_{i_2},j_{i_2}'}\cdots\delta_{j_{i_k},j_{i_k}'}$ by the $k$-uniformity of $\ket{\psi}$.
	
	Alice uses the projector $\ket{i}\bra{i}$ acting on the first system, $0\leq i\leq d_1-1$, then we can obtain
	\begin{equation}\ket{i,\psi_i}:=\sqrt{d_1}\sum_{(j_{i_2},\ldots,j_{i_k})\in\bbZ_{d_{i_2}}\times\ldots\times\bbZ_{d_{i_k}}}c_{i,j_{i_2},\ldots,j_{i_k}}\ket{i,j_{i_2},\ldots,j_{i_k}}\ket{\psi(i,j_{i_2},\ldots,j_{i_k})}
	\end{equation}
	with  probability $\frac{1}{d_1}$, where $\ket{\psi_i}\in \bbC^{d_{2}}\otimes\cdots\otimes\bbC^{d_{N}}$. Since
	\begin{equation}
	\begin{split}
	&(\ketbra{\psi_i}{\psi_i})_{\{i_2,i_3,\ldots, i_k\}}=\tr_{\{i_2,\ldots,i_k\}^c}\ketbra{\psi_i}{\psi_i}=\tr_{\{i_2,\ldots,i_k\}^c}d_1\sum_{(j_{i_2},\ldots,j_{i_k})\in\bbZ_{d_{i_2}}\times\cdots\times\bbZ_{d_{i_k}}}\sum_{(j_{i_2}',\ldots,j_{i_k}')\in\bbZ_{d_{i_2}}\times\cdots\times\bbZ_{d_{i_k}}}\\
	&c_{i,j_{i_2},\ldots,j_{i_k}}\overline{c_{i,j_{i_2}',\ldots,j_{i_k}'}}\ket{j_{i_2},\ldots,j_{i_k}}\bra{j_{i_2}',\ldots,j_{i_k}'}\otimes\ket{\psi(i,j_{i_2},\ldots,j_{i_k})}\bra{\psi(i,j_{i_2}',\ldots,j_{i_k}')}=\frac{1}{d_{i_2}\cdots d_{i_k}}I_{d_{i_2}\cdots d_{i_k}},	
	\end{split}
	\end{equation}
	then $\ket{\psi_i}$ is a $(k-1)$-uniform state in $\bbC^{d_{2}}\otimes\cdots\otimes\bbC^{d_{N}}$. Moreover, $\braket{\psi_i}{\psi_i'}=\delta_{i,i'}$ due to $\braket{\psi(i,j_{i_2},\ldots,j_{i_k})}{\psi(i',j_{i_2}',\ldots,j_{i_k}')}=\delta_{i,i'}\delta_{j_{i_2},j_{i_2}'}\cdots\delta_{j_{i_k},j_{i_k}'}.$
\end{proof}

\section{The proof of Proposition~\ref{Pro:k-basis}}
\label{Appendix:B}
\begin{proof}	
	Assume $\ket{\psi}$ is a $k$-uniform state with minimum support in $\bbC^{d_1}\otimes\bbC^{d_2}\otimes\cdots\otimes\bbC^{d_N}$, then $\ket{\psi}$ can be written as
	\begin{equation}
	\ket{\psi}=\sum_{(j_1,j_2,\ldots,j_k)\in \bbZ_{d_1}\times\bbZ_{d_2}\times\cdots\times\bbZ_{d_k}}c_{j_1,j_2,\ldots,j_k}\ket{j_1,j_2,\ldots,j_k}\ket{\psi(j_1,j_2,\ldots,j_k)},
	\end{equation}
	where $|c_{j_1,j_2,\ldots,j_k}|=\frac{1}{\sqrt{d_1d_2\cdots d_k}}$, $\ket{\psi(j_1,j_2,\ldots,j_k)}=\otimes_{\ell=k+1}^{N}\ket{m_\ell(j_1,j_2,\ldots,j_k)}$.

	Since local unitary operations do not change the $k$-uniformity, we only need to show the orthogonality of these states. We have
	\begin{equation}
	\begin{split}
	U(\bv)\ket{\psi}=&\sum_{(j_1,j_2,\ldots,j_k)\in \bbZ_{d_1}\times\bbZ_{d_2}\times\cdots\times\bbZ_{d_k}}c_{j_1,j_2,\ldots,j_k}(\otimes_{\ell=1}^kw_\ell^{j_\ell v_\ell}\ket{j_\ell})\\
	&\otimes[\otimes_{\ell=k+1}^N\ket{m_\ell(j_1,j_2,\ldots,j_k)\oplus_\ell v_\ell}].
	\end{split}
	\end{equation}
	Assume $\bv\neq\bu\in\bbZ_{d_1}\times\bbZ_{d_2}\times\cdots\times\bbZ_{d_N}$, and $\ell_0$ is the largest number such that $v_{\ell_0}\neq u_{\ell_0}$. We have
	\begin{equation}
\begin{split}
	\bra{\psi}U(\bu)^{\dagger}U(\bv)\ket{\psi}=&\sum_{(j_1,j_2,\ldots,j_k)\in \bbZ_{d_1}\times\bbZ_{d_2}\times\cdots\times\bbZ_{d_k}}\sum_{(j_1',j_2',\ldots,j_k')\in \bbZ_{d_1}\times\bbZ_{d_2}\times\cdots\times\bbZ_{d_k}}
	c_{j_1,j_2,\ldots,j_k}\overline{c_{j_1',j_2',\ldots,j_k'}}\\&(\prod_{\ell=1}^kw_\ell^{j_\ell v_\ell-j_\ell'u_\ell}\braket{j_\ell'}{j_\ell})\prod_{\ell=k+1}^{N}\braket{m_\ell(j_1',j_2',\ldots,j_k')\oplus_\ell u_\ell}{m_\ell(j_1,j_2,\ldots,j_k)\oplus_\ell v_\ell}\\
	=&\frac{1}{d_1d_2\cdots d_k}\sum_{(j_1,j_2,\ldots,j_k)\in \bbZ_{d_1}\times\bbZ_{d_2}\times\cdots\times\bbZ_{d_k}}(\prod_{\ell=1}^kw_\ell^{j_\ell (v_\ell-u_\ell)})\\
	&\prod_{\ell=k+1}^{N}\braket{m_\ell(j_1,j_2,\ldots,j_k)\oplus_\ell u_\ell}{m_\ell(j_1,j_2,\ldots,j_k)\oplus_\ell v_\ell}.
	\end{split}
\end{equation}
	If $\ell_0\geq k+1$, then $\braket{m_{\ell_0}(j_1,j_2,\ldots,j_k)\oplus_{\ell_0} u_{\ell_0}}{m_{\ell_0}(j_1,j_2,\ldots,j_k)\oplus_{\ell_0} v_{\ell_0}}=0$, $\bra{\psi}U(\bu)^{\dagger}U(\bv)\ket{\psi}=0$; If $\ell_0\leq k$, then
	\begin{equation}
\begin{split}
	\bra{\psi}U(\bu)^{\dagger}U(\bv)\ket{\psi}&=\frac{1}{d_1d_2\cdots d_k}\sum_{(j_1,j_2,\ldots,j_k)\in \bbZ_{d_1}\times\bbZ_{d_2}\times\cdots\times\bbZ_{d_k}}(\prod_{\ell=1}^kw_\ell^{j_\ell (v_\ell-u_\ell)})\\
	&=\frac{1}{d_1d_2\cdots d_k}\prod_{\ell=1}^{k}(\sum_{j_\ell=0}^{d_l-1}w_\ell^{j_\ell(v_\ell-u_\ell)})=0,
	\end{split}
\end{equation}
	due to $\sum_{j_{\ell_0}=0}^{d_{\ell_0}-1}w_{\ell_0}^{j_{\ell_0}(v_{\ell_0}-u_{\ell_0})}=0$.
\end{proof}

\section{The proof of Proposition~\ref{pro:local}}
\label{Appendix:C}
\begin{proof}
    We know that
	\begin{equation}
	\cB=\{U(\bv)\ket{\psi}:\bv\in\bbZ_4\times\bbZ_2\times\bbZ_2\times\bbZ_2\times\bbZ_2 \}
	\end{equation}
	is an orthogonal basis consisting of $2$-uniform states with minimum support in $\bbC^4\otimes(\bbC^2)^{\otimes 4}$,
	where $\ket{\psi}_{ABCDE}=\frac{1}{2\sqrt{2}}(\ket{00000}+\ket{01111}+\ket{10011}+\ket{11100}+\ket{20101}+\ket{21010}+\ket{30110}+\ket{31001})$. Assume Alice goes first, and starts a nontrivial orthogonality-preserving local measurement
	\begin{equation}
	E_m=M_m^{\dagger}M_m=
	\begin{bmatrix}
	a_{00} &a_{01}  &a_{02}  &a_{03}\\	
	a_{10} &a_{11}  &a_{12}  &a_{13}\\	
	a_{20} &a_{21}  &a_{22}  &a_{23}\\	
	a_{30} &a_{31}  &a_{32}  &a_{33}
	\end{bmatrix}.
	\end{equation}
	The postmeasurement states $\{M_m\otimes I_{BCDE} (U(\bv)\ket{\psi}):\bv\in\bbZ_4\times\bbZ_2\times\bbZ_2\times\bbZ_2\times\bbZ_2 \}$
	should be mutually orthogonal. Denote $N(\bv)=M_m\otimes I_{BCDE}(U(\bv)\ket{\psi})$.
	By Proposition~\ref{pro:projective}, we have $\ket{\psi}=\ket{0}_A\ket{\psi_0}+\ket{1}_A\ket{\psi_1}+\ket{2}_A\ket{\psi_2}+\ket{3}_A\ket{\psi_3}$, where $\braket{\psi_i}{\psi_j}=0$ for $0\leq i\neq j\leq 3$.
	Since $N((0,0,0,0,0))$ is orthogonal to $\{N(\bv):\bv=\{(1,0,0,0,0),(2,0,0,0,0),(3,0,0,0,0)\}\}$, we have
	\begin{equation}
	\left\{\begin{split}
	&a_{00}+w_4a_{11}+w_4^2a_{22}+w_4^3a_{33}=0,\\
	&a_{00}+w_4^2a_{11}+w_4^4a_{22}+w_4^6a_{33}=0,\\
	&a_{00}+w_4^3a_{11}+w_4^6a_{22}+w_4^9a_{33}=0.
	\end{split}
	\right.
	\end{equation}
	Then, we obtain $a_{00}=a_{11}=a_{22}=a_{33}$. Since $N((0,0,0,0,0))$ is orthogonal to $\{N(\bv):\bv=\{(0,0,0,1,1),(1,0,0,1,1),(2,0,0,1,1),(3,0,0,1,1)\}\}$,  $\{N(\bv):\bv={(0,0,1,0,1),(1,0,1,0,1),(2,0,1,0,1),(3,0,1,0,1)\}\}}$ and $\{N(\bv) :\bv={(0,0,1,1,0),(1,0,1,1,0),(2,0,1,1,0),(3,0,1,1,0)\}\}}$, we obtain $a_{01}=a_{10}=a_{23}=a_{32}=0$, $a_{02}=a_{20}=a_{13}=a_{31}=0$ and $a_{03}=a_{30}=a_{12}=a_{21}=0$. Thus, $E_m$ is proportional to the identity operator, and it means that Alice can not go first.
	
	Assume Bob goes first and starts
	\begin{equation}
	E_m=M_m^{\dagger}M_m=
	\begin{bmatrix}
	b_{00} &b_{01} \\	
	b_{10} &b_{11}
	\end{bmatrix}.
	\end{equation}
	Denote $N(\bv)=I_A\otimes M_m\otimes I_{CDE} (U(\bv)\ket{\psi})$. Since $N((0,0,0,0,0))$ is orthogonal to  $\{N(\bv):\bv=\{(0,1,0,0,0)，(0,0,1,1,1),(0,1,1,1,1)\}\}$, we obtain $b_{00}=b_{11}$ and $b_{01}=b_{10}=0$. Thus Bob can not go first.
	
	Assume Charles goes first and starts
	\begin{equation}
	E_m=M_m^{\dagger}M_m=
	\begin{bmatrix}
	c_{00} &c_{01} \\	
	c_{10} &c_{11}
	\end{bmatrix}.
	\end{equation}
	Denote $N(\bv)=I_{AB}\otimes M_m\otimes I_{DE} (U(\bv)\ket{\psi})$. Since $N((1,0,0,0,0))$ is orthogonal to $\{N(\bv):\bv=\{(0,1,0,0,0),(0,1,1,0,0)\}\}$, and $N((0,0,0,0,0))$ is orthogonal to $\{N(\bv):\bv=\{(0,0,1,0,0)\}\}$, we obtain $c_{00}=c_{11}$ and $c_{01}=c_{10}=0$. Thus Charles can not go first.
	
	Assume David goes first and starts
	\begin{equation}
	E_m=M_m^{\dagger}M_m=
	\begin{bmatrix}
	d_{00} &d_{01} \\	
	d_{10} &d_{11}
	\end{bmatrix}.
	\end{equation}
	Denote $N(\bv)=I_{ABC}\otimes M_m\otimes I_{E} (U(\bv)\ket{\psi})$. Since $N((2,0,0,0,0))$ is orthogonal to $\{N(\bv):\bv=\{(0,1,0,0,0),(0,1,0,1,0)\}\}$, and $N((0,0,0,0,0))$ is orthogonal to $\{N(\bv):\bv=\{(0,0,0,1,0)\}\}$, we obtain $d_{00}=d_{11}$ and $d_{01}=d_{10}=0$. Thus David can not go first.
	
	Assume Eve goes first and starts
	\begin{equation}
	E_m=M_m^{\dagger}M_m=
	\begin{bmatrix}
	e_{00} &e_{01} \\	
	e_{10} &e_{11}
	\end{bmatrix}.
	\end{equation}
	Denote $N(\bv)=I_{ABCD}\otimes M_m (U(\bv)\ket{\psi})$. Since $N((1,0,0,0,0))$ is orthogonal to $\{N(\bv):\bv=\{(0,1,0,0,0),(0,1,0,0,1)\}\}$, and $N((0,0,0,0,0))$ is orthogonal to $\{N(\bv):\bv=\{(0,0,0,0,1)\}\}$, we obtain $e_{00}=e_{11}$ and $e_{01}=e_{10}=0$. Thus Eve can not go first.
	
	In conclusion, $\cB$ is locally irreducible.
\end{proof}

\bibliographystyle{IEEEtran}
\bibliography{reference}
\end{document}